\documentclass[12pt]{article}
\usepackage{amsthm}
\usepackage{caption}
\usepackage{amssymb,setspace}
\usepackage[margin=1in,lines=25]{geometry}
\usepackage{amsmath}
\usepackage{natbib}
\usepackage[colorlinks,citecolor=blue,urlcolor=blue,filecolor=blue,backref=page]{hyperref}
\usepackage{graphicx}
\usepackage{amsmath,bm}
\usepackage{epstopdf}
\usepackage{mathtools}
\usepackage{ulem}
\usepackage{color}
\usepackage{subfigure,enumerate}
\usepackage{amsmath}
\usepackage{algorithm}
\usepackage[noend]{algpseudocode}
\usepackage{booktabs}
\usepackage{natbib}

\title{}
\author{}
\def\bfomega{\boldsymbol\omega}
\def\bbeta{\boldsymbol\beta}

\def\bbeta{\boldsymbol \beta}

\def\bfh{\mathbf t}
\def\mTGLG{\mathrm{TGLG}}

\newcommand{\cm}[1]{\ignorespaces}

\def\bfx{\mathbf x}

\def\bfz{\mathbf z}

\def\bfI{\mathbf I}

\def\bfL{\mathbf L}

\def\bfalpha{\boldsymbol \alpha}
\def\bfbeta{\boldsymbol \beta}

\def\bfgamma{\boldsymbol \gamma}

\def\bfmu{\boldsymbol\mu}
\def\bfSigma{\boldsymbol\Sigma}

\def\bfzero{\boldsymbol 0}

\def\cA{\mathcal A}

\def\go{\rightarrow}

\def\mE{\mathrm{E}}

\def\mPr{\mathrm{Pr}}

\def\mP{\mathrm{P}}

\def\mN{\mathrm{N}}

\def\mIG{\mathrm{IG}}

\def\mPr{\mathrm{Pr}}
\def\mE{\mathrm{E}}

\def\bbeta{\boldsymbol \beta}

\def\bgamma{\boldsymbol \gamma}

\def\cA{\mathcal{A}}

\def\rT{\mathrm T}

\def\go{\rightarrow}

\newtheorem{mythm}{Theorem}

\makeatletter
\g@addto@macro\normalsize{%
  \setlength\abovedisplayskip{0.3mm}
  \setlength\belowdisplayskip{0.3mm}
  \setlength\abovedisplayshortskip{1mm}
  \setlength\belowdisplayshortskip{1mm}
}
\makeatother

\title{Bayesian network marker selection via the thresholded graph Laplacian Gaussian prior}
\author{Qingpo Cai\thanks{Qingpo Cai is Ph.D. Student, Department of Biostatistics and Biostatistics, Emory University, Atlanta, GA 30322,  Jian Kang is Associate Professor, Department of Biostatistics, University of Michigan, Ann Arbor, MI 48109.  Tianwei Yu is Associate Professor, Department of Biostatistics and Bioinformatics, Emory University, Atlanta, GA 30322.},\quad   Jian Kang,\quad Tianwei Yu\thanks{To whom correspondence should be addressed: jiankang@umich.edu or tianwei.yu@emory.edu}}
\providecommand{\keywords}[1]{\textbf{\textit{keywords:}} #1}
\bibpunct{(}{)}{,}{a}{}{;}

\begin{document}
\maketitle
\thispagestyle{empty}
\begin{abstract}
\setlength{\baselineskip}{18pt}	
Selecting informative nodes over large-scale networks  becomes increasingly important in many research areas.  Most existing methods focus on the local network structure and incur heavy computational costs for the large-scale problem.  In this work, we propose a novel prior model for Bayesian network marker selection in the generalized linear model (GLM) framework: the Thresholded Graph Laplacian Gaussian (TGLG) prior, which adopts the graph Laplacian matrix to characterize the conditional dependence between neighboring markers accounting for the global network structure. Under mild conditions, we show the proposed model enjoys the posterior consistency with a diverging number of edges and nodes in the network. We also develop a Metropolis-adjusted Langevin algorithm (MALA) for efficient posterior computation, which is scalable to large-scale networks. We illustrate the superiorities of the proposed method compared with existing alternatives via extensive simulation studies and an analysis of the breast cancer gene expression dataset in the Cancer Genome Atlas (TCGA).

\bigskip
\noindent \keywords{Bayesian Variable Selection, Gene Network,  Thresholded Graph Laplacian Gaussian Prior,  Generalized Linear Model, Posterior Consistency}
\end{abstract}

%\begin{keywords}

%\end{keywords}
\newpage
\setcounter{page}{1}

\section{Introduction}
\textwidth 6in
\textheight 8.25in
 In biomedical research,  complex biological systems are often modeled or represented as biological networks~\citep{kitano2002systems}. High-throughput technology such as next generation sequencing~\citep{schuster2007next}, mass spectrometry~\citep{aebersold2003mass} and medical imaging~\citep{doi2007computer} has generated massive datasets related to those biological networks.  For example, in omics studies, a biological network may represent the interactions or dependences among a large set of genes/proteins/metabolites; and the expression data are a number of observations at each node of the network~\citep{barabasi2011network}. In neuroimaging studies, a biological network may refer to the functional connectivity among many brain regions or voxels; and the neural activity can be measured at each node of the network.   In many biomedical studies, one important research question is to select informative nodes from tens of thousands of candidate nodes that are strongly associated with the disease risk or other clinical outcomes~\citep{greicius2003functional}. We refer to these informative nodes as network markers~\citep{kim2012multi, peng2014brain, yuan2017network} and the selection procedure as network marker selection. One promising solution is to perform network marker selection under regression framework where the response variable is the clinical outcome and predictors are nodes in the network. The classical variable selection~\citep{george1993variable,fan2001variable} in the regression model can be considered as a special case of the network marker selection, where the variable refers to the nodes in a network without edges.

For variable selection in regression models, many regularization methods have been proposed with various penalty terms, including the least absolute shrinkage and selection operator or the $L_1$ penalty~\citep[LASSO]{tibshirani1996regression,zou2006adaptive},  elastic-net or the $L_1$ plus $L_2$ penalty~\citep{zou2005regularization},  the Smoothly Clipped Absolute Deviation penalty~\citep[SCAD]{ fan2001variable}, the minimax concave penalty~\citep[MCP]{zhang2010nearly} and so on. Several network constrained regularization regression approaches have been developed to improve selection accuracy and increase  prediction power. One pioneering work is the graph-constrained estimation~\citep[Grace]{li2008network}, which adopts the normalized graph Laplacian matrix to incorporate the network dependent structure between connected nodes.  As an extension of Grace, the adaptive Grace~\citep[aGrace]{li2010variable} makes constraints on the absolute values of weighted coefficients between connected nodes. Alternatively,  an $L_\gamma$ norm group penalty~\citep{pan2010incorporating}  and a fused LASSO type penalty~\citep{luo2012two} have been proposed  to penalize the difference of absolute values of coefficients between neighboring nodes.  Instead of imposing constraints on coefficients between neighboring nodes, an $L_0$ loss to penalize their selection indicators~\citep{kim2013network} has been proposed, leading to a non-convex optimization problem for parameter estimation, which can be solved by approximating the non-continuous $L_0$ loss using the truncated lasso penalty (TLP).

%Genes in the biological system function in a tightly controlled network. Proximity of genes on the network has functional implications such as co-regulation or interaction etc. Taking into account the network information between nodes can result in more robust and interpretable results.

In addition to frequentist approaches, Bayesian variable selection methods have received much attention recently with many successful applications. The Bayesian methods are natural to  incorporate the prior knowledge and make  posterior inference on uncertainty of variable selection. A variety of prior models have been studied, such as the spike and slab prior~\citep{george1993variable}, the LASSO prior \citep{park2008bayesian}, the Horseshoe prior~\citep{polson2012local}, the non-local prior~\citep{johnson2012bayesian} , the Dirichlet Laplace prior \citep{bhattacharya2015dirichlet} and more. To incorporate the known network information, \cite{stingo2011incorporating} employed a Markov random field to capture network dependence and jointly select pathways and genes;  and \cite{chekouo2016bayesian} adopted a similar approach for imaging genetics analysis. \cite{zhou2013bayesian} proposed rGrace, a Bayesian random graph-constrained model to combine network information with empirical evidence for pathway analysis. A partial least squares (PLS) g-prior was developed in \cite{peng2013integrative} to incorporate prior knowledge on gene-gene interactions or functional relationship for identifying genes and pathways. \cite{chang2016scalable} proposed a Bayesian shrinkage prior which smoothed shrinkage parameters of connected nodes to a similar degree for structural variable selection.

The Ising model is another commonly used Bayesian structural variable selection method. It has been used as a prior model for latent selection indicators that lay on an undirected graph characterizing the local network structure. They are especially successful  for variable selection over the grid network motivated by some applications, for example, the motif finding problem~\citep{doi:10.1198/jasa.2010.tm08177}  and the imaging data analysis~\citep{goldsmith2014smooth, li2015spatial}. However, it is very challenging for fully Bayesian inference on the Ising model over the large-scale network due to at least two reasons: 1) The posterior inference can be quite sensitive to the hyperparameter specifications in the Ising priors based on empirical Bayes estimates or subjective prior elicitation in some applications. However, fully Bayesian inference on those parameters is difficult due to the intractable normalizing constant in the model.   2) Most posterior computation algorithms, such as the single-site Gibbs sampler and the Swendsen-Wang algorithm, incur heavy computational costs for updating the massive binary indicators over large-scale networks with complex structures. {\color{blue} In addition, \cite{dobra2009variable, kundu2015bayesian, liu2014bayesian} and \cite{peterson2016joint} also proposed Bayesian variable selection approaches for predictors with unknown network structure.}

%%  {\color{red}$<$Does this refer to the bottom of page 3? should we move them here?$>$} {\color{blue} I think Tianwei refers to the blue part above. Do we need to move it here?}
%
% {\color{blue}Some hierarchical or multilevel variable selection methods~\citep{zhang2014bayesian, zhaobayesian2016} have been proposed by utilizing additional prior knowledge information to improve the selection accuracy. In particular, for the gene and the sub-network selection problem, some Bayesian methods have been developed for jointly selecting genes and pathways~\citep{stingo2011incorporating, zhe2013joint} by using two levels of selection indictors that incorporate pathway information along with the existing gene networks.}
%  Although the pathway information is helpful to facilitate biologically meaningful gene selection and reduce the posterior computation cost using the hierarchical selection approach, it is not always available for specific applications. In addition, the delineation of pathways can be artificial, and irrelevant or inaccurate pathway information may mislead  the model inference. Thus, there are needs of developing computationally efficient Bayesian network marker selection methods for large-scale networks without using the extra layer of pathway information.

To address limitations of existing methods,  we propose a new prior model: the thresholded graph Laplacian Gaussian (TGLG) prior, to perform network marker selection over the large-scale network by thresholding a latent continuous variable attached to each node. To model the selection dependence over the network, all the latent variables are assumed to follow a multivariate Gaussian distribution with mean zero and covariance matrix constructed by a normalized graph Laplacian matrix. The effect size of each node is modeled through an independent Gaussian distribution.

Threshold priors have been proposed for Bayesian modeling of sparsity in various applications.  Motivated by the analysis of financial time series data, \cite{nakajima2013bayesian_a} and \cite{nakajima2013bayesian_b} proposed a latent threshold approach to imposing dynamic sparsity in the simultaneous autoregressive models (SAR). \cite{nakajima2017dynamics} further extended this type of models for the analysis of EEG data.  To analyze neuroimaging data, \cite{shi2015thresholded} proposed a hard-thresholded Gaussian process prior for image-on-scalar regression; and \cite{kang2018scalar} introduced a soft-thresholded Gaussian process for scalar-on-image regression. To construct the directed graphs in genomics applications, \cite{ni2017bayesian} adopted a hard threshold Gaussian prior in a structural equation model.  However, all the existing threshold prior models do not incorporate the useful network structural information, and thus are not directly applicable to the network marker selection problem of our primary interest.

In this work, we propose to build the threshold priors using the graph Laplacian matrix,  which has been used to capture the structure dependence between neighboring nodes \citep{li2008network, zhe2013joint, li2010variable}. Most of those frequentist methods directly specify the graph Laplacian matrix from the existing biological network. %
\cite{liu2014bayesian} has proposed a Bayesian regularization graph Laplacian (BRGL) approach which utilizes the graph Laplacian matrix  to specify {\it a priori} precision matrix of regression coefficients. However, BRGL is fundamentally different from our method in that it is one type of continuous shrinkage priors for regression coefficients which have quite different prior supports compared with our TGLG priors.  BRGL were developed only for linear regression and its computational cost can be extremely heavy for large-scale networks. In addition, there is lack of theoretical justifications for BRGL when the large-scale network has a diverging number of edges and nodes.

Our method is a compelling Bayesian approach to network marker selection.  The TGLG prior has at least four markable features: 1) Fully Bayesian inference for large-scale networks is feasible in that the TGLG prior does not involve any intractable normalizing constants. 2) Posterior computation can be more efficient,   since the TGLG-based inference avoids updating the latent binary selection indicators and instead updates the latent continuous variables, to which many existing approximation techniques can be potentially applied. 3) The graph Laplacian matrix~\citep{chung1997spectral, li2008network, zhe2013joint} based prior can incorporate the topological structure of the network which has been adopted in genomics. 4) The TGLG prior enjoys the large support for Bayesian network marker selection over large-scale networks, leading to  posterior consistency of model inference with a diverging number of nodes and edges under the generalized linear model (GLM) framework.

The remainder of the manuscript is organized as follows. In Section 2, we introduce the TGLG prior and propose our model for network marker selection under the GLM framework. In Section 3, we study the theoretical properties for the TGLG prior and show the posterior consistency of model inference. In Section 4, we discuss the hyper prior specifications and the efficient posterior computation algorithm. We illustrate the performance of our approach via simulation studies and an application on the breast cancer gene expression dataset from The Cancer Genome Atlas (TCGA) in Section 5. We conclude our paper with a brief discussion on the future work in Section 6.

\section{The Model}
\label{sec2}
Suppose the observed dataset includes a network with $p_n$ nodes, one response variable and $q$ confounding variables. For each node, we have $n$ observations. For observation $i$, $i = 1,\ldots, n$, let $y_i$ be the response variable, $\bfx_i = (x_{i1}, \cdots, x_{ip_n})^{\rT}$ be the vector of nodes and $\bfz_i = (z_{i1}, \cdots, z_{iq})^{\rT}$ be the vector of confounding variables. Denote by $D_n = \{\bfz_i, \bfx_i, y_i\}^n_{i=1}$ the dataset. We write the number of nodes as $p_n$ to emphasize on the diverging number of nodes in our asymptotical theory. Drop subscript $i$ to have generic notation for a response variable $y$, a vector of nodes $\bfx$ and a vector of confounders $\bfz$. Generalized linear model (GLM) is a flexible regression model to relate a response variable to a vector of nodes and confounding variables. The GLM density function for $(y, \bfx, \bfz)$ with one natural parameter is:
\begin{equation}\label{eq:glm}
f^*(y, h^*) = \exp \{a(h^*)y + b(h^*) + c(y)\},
\end{equation}
where $h^* = \bfz^{\rT} \bfomega^* + \bfx^{\rT} \bbeta^*$ is the linear parameter in the model, $\bfomega^*$ and $\bbeta^*$ are true coefficients that generate data, $a(h)$ and $b(h)$ are continuous differentiable functions. The true mean function is
\begin{equation*}
\mu^* = \mE(y\mid \bfz, \bfx) = -b' (h^*) / a'(h^*) \equiv g^{-1} (\bfz^{\rT} \bfomega^* + \bfx^{\rT} \bbeta^*),
\end{equation*} where $g^{-1}(\cdot)$ is an inverse link function, which can be chosen according to the specific type of the response variable. For example, one can choose the identity link for the continuous response and the logit link for the binary response.

\begin{figure}[htbp]
  \centering
 \subfigure[network]{
%\label{fig:subfig:a} %% label for first subfigure
\includegraphics[width=2in]{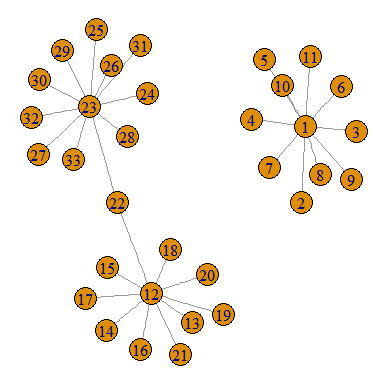}
}
\subfigure[correlation]{
%\label{fig:subfig:b} %% label for second subfigure
\includegraphics[width=2.5in, height=2.2in]{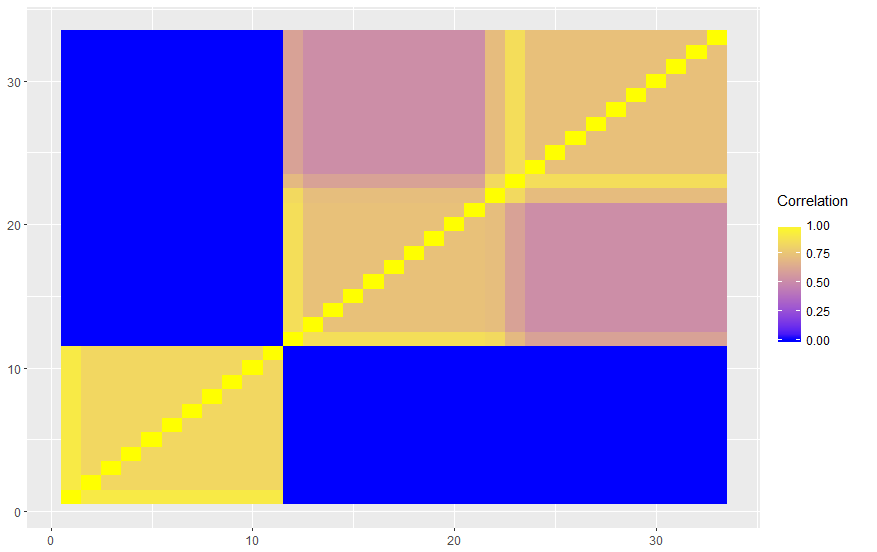}}

\caption{An example of the graph and the corresponding correlation matrix of $\bfgamma$ that was constructed from the inverse graph Laplacian matrix}
\label{exam1} %% label for entire figure
\end{figure}

In \eqref{eq:glm}, coefficient vector $\bfomega$ is a nuisance parameter to adjust for the confounder effects, for which we assign a Gaussian prior with mean zero and independent covariance, i.e. $\bfomega\sim \mN(0,\sigma^2_\omega\bfI_q)$ for $\sigma^2_\omega> 0$. Here $\bfI_d$ represents an identity matrix  of dimension $d$ for any $d>0$. Coefficient vector $\bfbeta$ represents the effects of nodes on the response variable. Here we perform network marker selection by imposing sparsity on $\bfbeta$. To achieve this goal, we develop a new prior model for $\bfbeta$: the thresholded graph Laplacian Gaussian (TGLG) prior.   Suppose the observed network can be represented by a graph $G$, with each vertex corresponding to one node in the network.  Let $j \sim k$ indicate there exists an edge between vertices $j$ and $k$ in $G$. Let $d_j$ represent the degree of vertex $j$, i.e., the number of nodes that are connected to vertex $j$ in $G$. Denote by  $\bfL = (L_{jk})$ a $p_n\times p_n$ normalized graph Laplacian matrix, where $L_{jk} = 1$ if $j=k$ and $\text{deg}(v_j)\neq 0$, $L_{jk} = -1/\sqrt{d_j d_k}$ if $j \sim k$, and $L_{jk} = 0$ otherwise. For any $d>0$, denote by $\bfzero_d$ an all zero vector of dimension $d$. For any $\lambda, \varepsilon, \sigma_{\alpha}^2, \sigma_\gamma^2 > 0$,  we consider an element-wise decomposition of $\bfbeta$ for the prior specifications:
\begin{equation} \label{eq:TGLG}
{\footnotesize
\begin{aligned}
\bfbeta = \bfalpha \circ \bfh_\lambda(\bfgamma), \quad\bfgamma \sim \mN\{\bfzero_{p_n}, \sigma_\gamma^2(\bfL + \varepsilon \bfI_{p_n})^{-1} \}, \quad \bfalpha \sim \mN(\bfzero_{p_n}, \sigma_{\alpha}^2\bfI_{p_n}).
\end{aligned}
}
\end{equation}
Here $\bfalpha = (\alpha_1,\ldots, \alpha_{p_n})^{\rT}$ represents the effect size of nodes. The operator "$\circ$" is the element-wise product. The vector thresholding function is $\bfh_\lambda(\bfgamma) = \{I(|\gamma_1| > \lambda), \ldots, I(|\gamma_{p_n}| > \lambda)\}^{\rT}$, where $I(\cA)$ is the event indicator with $I(\cA) = 1$ if $\cA$ occurs and $I(\cA) = 0$ otherwise.  The latent continuous vector $\bfgamma = (\gamma_1,\ldots, \gamma_{p_n})^{\rT}$ controls the sparsity over graph $G$.  We refer to \eqref{eq:TGLG} as the TGLG prior for $\bfbeta$,  denoted as $\bfbeta \sim \mTGLG(\lambda, \varepsilon,\sigma^2_\gamma,\sigma^2_\alpha)$.  The TGLG prior implies that for any two nodes $j$ and $k$,  $\gamma_j$ and $\gamma_k$ are conditionally dependent given others if and only if $j\sim k$ over the graph $G$. In this case, their absolute values are more likely to be smaller or larger than a threshold value $\lambda$ together. This further implies that nodes $j$ and $k$ are more likely to be selected as network marker or not selected together if $j\sim k$. Figure \ref{exam1} shows an example of a graph and the corresponding correlation matrix of $\bfgamma$ for $\varepsilon = 10^{-2}$, where the $\gamma$'s of connected vertices are highly correlated.

 There are four hyperparameters in the TGLG prior model. The threshold $\lambda$ controls {\it a priori} the sparsity. When $\lambda \to 0$, all the nodes tend to be selected. When $\lambda \to \infty$, none of them will be selected. The parameter $\varepsilon$ determines the impact of the network on the sparsity. When $\varepsilon \to \infty$, $\gamma$'s of connected vertices tend to be independent while they tend to be perfectly correlated when $\varepsilon \to 0$. The two variance parameters $\sigma^2_\gamma$ and $\sigma^2_\alpha$ control the prior variability of the latent vectors $\bfgamma$ and $\bfalpha$ respectively.

Now we discuss how to specify the hyperparameters.  For variance terms $\sigma^2_\gamma$ and $\sigma^2_\alpha$, we use the conjugate prior model by assigning the Inverse-Gamma distribution $\mIG(a_\gamma, b_\gamma)$ and $\mIG(a_\alpha, b_\alpha)$ respectively. We fix $\sigma^2_\omega$ as a large value. We assign the uniform prior to the threshold parameter $\lambda$, i.e. $\lambda \sim \text{Unif} (0, \lambda_u)$ with upper bound $\lambda_u>0$. We choose a wide range by set $\lambda_u = 10$ in the rest of manuscript. For parameter $\varepsilon$, we can either assign an log-normal prior ($\text{log}\varepsilon \sim N(\mu_\varepsilon, \sigma^2_\varepsilon)$) or set as a fixed small value.

\section{Theoretical Properties}
\label{theory}
In this section, we examine the theoretical properties of TGLG prior based network marker selection under the GLM framework. In particular, we establish the posterior consistency with a diverging number of nodes in the large-scale networks.

 Let $\xi \subset \{1, 2, \cdots, p_n \}$ denote the set of selected node indices, i.e. $I(|\gamma_j|>\lambda) = 1$, if $j \in \xi$, $I(|\gamma_j|>\lambda) = 0$, otherwise. The number of nodes in $\xi$ is denoted as $|\xi|$.  For a model $\xi = (i_1, \cdots, i_{|\xi|})$,  denote by  $\bbeta_\xi = (\beta_{i_1}, \cdots, \beta_{i_{|\xi|}})^{\rT}$ the coefficient of interest, respectively. Let $\pi( \xi, d \bbeta_\xi, d \bfomega)$ represent the joint prior probability measure for model $\xi$, parameters $\bbeta_\xi$ and confounding coefficients $\bfomega$. Their joint posterior probability measure conditional on dataset $D_n$ is:
$$\pi( \xi, d \bbeta_\xi, d \bfomega \mid D_n) = \frac{\prod_{i=1}^n f(y_i, h_i) \pi( \xi, d \bbeta_\xi, d \bfomega)}{ \sum_{\xi'} \int_{\bbeta_{\xi'}} \prod_{i=1}^n f(y_i, h_i) \pi( \xi', d \bbeta_{\xi'}, d \bfomega)},$$
where  $f(y_i, h_i) = \exp \{a(h_i)y_i + b(h_i) + c(y_i)\}$ be the density function of $y_i$ given $\bfx_i$ and $\bfz_i$ based on GLM with $h_i = \bfz_i^{\rT} \bfomega +\bfx^{\rT}_i \bbeta$. We examine  asymptotic properties of the posterior distribution of the density function $f$ regarding to the Hellinger distance~\citep{jiang2007bayesian,song2015split} under some regularity conditions.  The Hellinger distance $d(f_1, f_2)$ between two density functions $f_1(x,y)$ and $f_2(x,y)$ is defined as $$d(f_1, f_2) = \left[\int \int \{f_1^{1/2}(x,y) - f_2^{1/2}(x, y)\} dx dy\right]^{1/2}.$$ We list all the regularity conditions in the Appendix.
 We show that the TGLG prior and the proposed model enjoy the following properties:
\begin{mythm}\label{thm:1} (Large Support for Network Marker Selection)
 Assume a sequence $\epsilon_n \in (0, 1]$ with $n \epsilon^2_n \to \infty$ and a sequence of nonempty models $\xi_n$. Assume conditions (C1)--(C3) and (C7) hold. Given $\sigma^2_\alpha$ and $\sigma^2_\gamma$, for any sufficiently small $\eta > 0$, there exists $N_\eta$ such that for all $n > N_\eta$, we have
\begin{equation} \label{con11}
\pi(\xi = \xi_n ) \geq e^{-n \epsilon_n^2 / 128} \mbox{ and }
\end{equation}
\begin{equation} \label{con12}
\pi( \bfbeta_\xi \in B(\xi_n, \eta) \mid \xi = \xi_n) \geq e^{-n \epsilon_n^2 / 128} \mbox{ with } B(\xi_n, \eta) = \{\beta^*_j \pm \eta \epsilon_n^2 / |\xi_n|\}_{j \in \xi_n}.
\end{equation}
There exists $C_n > 0$, such that for all sufficiently large $n$ and for any $j \in \xi_n$:
\begin{equation} \label{con13}
\pi( |\beta_j| > C_n \mid \xi_n ) \leq e^{-n \epsilon_n^2 / 4}.
\end{equation}
\end{mythm}
\noindent This theorem shows that the TGLG prior has a large support for the network marker selection. Particularly, \eqref{con11} states that the TGLG prior  can select the true network marker with a positive prior probability bounded away from zero, \eqref{con12} ensures that the prior probability of the coefficients falling within an arbitrarily small neighborhood of the true coefficients with probability bounded away from zero, and \eqref{con13} indicates a sufficiently small tail probability of the TGLG prior.

\begin{mythm}\label{thm:2}
(Posterior Consistency for Network Marker Selection) \quad For the GLM with bounded covariates, i.e. $|x_j| \leq M$ for all $j=1, \cdots, p_n$ and $|z_k| \leq M$ for all $k=1, \cdots, q$, suppose the true node regression coefficients satisfy $$\lim_{n \to \infty} \sum_{j=1}^{p_n} |\beta^*_j| < \infty.$$ Let $\epsilon_n \in (0, 1]$ be a sequence such that $n \epsilon_n^2 \to \infty$. Assume conditions (C1)--(C7) hold.
Then we have the following results:

\begin{itemize}
\item[(i)]  Posterior consistency:
\begin{equation}\label{eq:post_const}
\lim_{n \to \infty} \mP\{\pi [ d(f, f^*) \leq \epsilon_n | D_n] \geq 1 - 2 e^{-n \epsilon_n^2 / 64}\} =1,
\end{equation}
where $f$ is the density function sampled from the posterior distribution and $f^*$ is the true density function.
\item[(ii)] For all sufficiently large $n$:
\begin{equation}\label{eq:prob_converge_rate}
\mP\{\pi [ d(f, f^*) >  \epsilon_n | D_n] \geq  2 e^{-n \epsilon_n^2 / 64}\} \leq 2 e^{-n \epsilon_n^2 / 64}.
\end{equation}
\item[(iii)] For all sufficiently large $n$:
\begin{equation}\label{eq:e_converge_rate}
\mE \{\pi [ d(f, f^*) > \epsilon_n | D_n] \} \leq 4 e^{-n \epsilon_n^2 / 32},
\end{equation}
Probability measure $\mP$  and expectation $\mE$ are both with respect to data $D_n$ that are generated from the true density $f^*$.
\end{itemize}

\end{mythm}
\noindent This theorem establishes the posterior consistency of network marker selection. In particular, \eqref{eq:post_const} implies that the posterior distribution of density $f$ concentrates on an arbitrarily small neighborhood of the true density $f^*$ under the Hellinger distance with a large probability. This probability converges to one as sample size $n\go \infty$. \eqref{eq:prob_converge_rate} provides the convergence rate of the posterior distribution indicating how fast the tail probability approaches to zero. \eqref{eq:e_converge_rate} indicates the average convergence rate of the posterior distribution of density $f$ concentrating on the arbitrarily small neighborhood of the true density $f^*$.

Please refer to the Supplementary File 1  for proofs of Theorems \ref{thm:1} and \ref{thm:2}.

\section{Posterior Computation}
\label{sec4}
Our primary goal is to make posterior inference on regression coefficients for network markers, i.e. $\bbeta$. According to the model specification, the sparsity of  $\beta_j$ is determined by the sparsity of $\alpha_j$ and whether $|\gamma_j|$ is less than $\lambda$ or not, i.e. $I(\beta_j = 0) = I(\alpha_j=0) I(|\gamma_j|\leq \lambda)$.  Since $\alpha_j$ has a non-sparse normal prior,  the posterior inclusion probability of node $j$ is just equal to  the posterior probability of $|\gamma_j|$ being greater than  $\lambda$; and given $\beta_j \neq 0$, the effect-size can be estimated by $E(\beta_j ||\gamma_j| > \lambda, D_n)$. All other parameters in the model can be estimated by its posterior expectations.

%\subsection{Metropolis-adjusted Langevin algorithm}
To simulate the joint posterior distribution for all parameters,   we adopt an efficient Metropolis-adjusted Langevin algorithm (MALA)~\citep{roberts1998optimal} for posterior computation. We introduce a smooth approximation for the thresholding function:
$$I(|\gamma_j|>\lambda) \simeq \frac{1}{2}\left\{1+\frac{2}{\pi}\text{arctan}\left(\frac{\gamma_j^2 - \lambda^2}{\varepsilon_0}\right)\right\} \quad  \text{for} \quad \varepsilon_0 \to 0,$$
leading to the analytically tractable first derivative: $$\frac{\partial \beta_j} {\partial \gamma_j} = \alpha_j \frac{2 \gamma_j/ \varepsilon_0}{\pi(1 + (\gamma_j^2 - \lambda^2)^2 / \varepsilon^2_0)}.$$
We choose $\varepsilon_0 = 10^{-8}$ in the simulation studies and real data application in this article.
%We use MALA to update $\alpha$ and $\bgamma$, random walk to update $\lambda$ with a truncated normal proposal distribution and update $\bfomega$ using a normal proposal distribution, Gibbs sampling for $\sigma_\alpha^2$, $\sigma_\gamma^2$.

%We choose the latter one with $\varepsilon = 10^{-5}$ for our simulation studies and real data analysis achieving the satisfactory performance.

Denote by $f(y_i \mid \bfomega, \bfalpha, \bgamma, \lambda)$ the likelihood function for all the parameters of interest for observation $i$. Let $\phi(\bfx\mid \bfmu,\bfSigma)$ denote the density function of a multivariate normal distribution with mean $\bfmu$ and covariance matrix $\bfSigma$ and $\phi_{+}(x \mid \mu,  \mu_l, \mu_u, \sigma^2)$ denote the density of a truncated normal distribution $N_{+} (\mu,  \mu_l, \mu_u, \sigma^2)$ density with mean $\mu$, variance $\sigma^2$ and interval $[\mu_l, \mu_u]$. Let $V_\omega = \sigma^2_\omega I_q$ be the variance of the prior distribution for $\bfomega$. Let $\Lambda_\gamma = (\bfL + \varepsilon \bfI_{p_n})^{-1}$.
The key steps in our posterior computation algorithm include:
\begin{itemize}
\item Update $\bfomega$ (Random Walk):  \quad Given current $\bfomega$, Draw $\bfomega^{new} \sim\mN(\bfomega, \tau^2_\omega I_q)$. Set $\bfomega  \longleftarrow \bfomega^{new}$ with probability min$\left\{1, \frac{ \phi( \bfomega^{new} \mid 0, V_\omega) \prod_i f(y_i \mid \bfomega^{new}, \bullet) }{\phi(\bfomega \mid 0, V_\omega) \prod_i f(y_i \mid \bfomega, \bullet) }\right\}$.
\item Update $\bgamma$ (MALA): \quad Given current $\bgamma$, draw $\bgamma^{new} \sim\mN\{ \mu(\bgamma), \tau^2_\gamma I_p\}$, where $\mu(\bgamma) = \bgamma+ \frac{\tau^2_\gamma}{2} ( \frac{\partial \text{log}f}{\partial \bgamma}- \frac{1}{2} \sigma_\gamma^2 \Lambda_\gamma \bgamma)$ with $\frac{\partial \text{log}f}{\partial \gamma_j} = \sum_{i=1}^n (a' (\bfz^{\rT}_i \bfomega + \bfx^{\rT}_i \bbeta) + b' (\bfz^{\rT}_i \bfomega + \bfx^{\rT}_i \bbeta)) x_{ij}\frac{\partial \beta_j} {\partial \gamma_j}$. Set $\bgamma  \longleftarrow \bgamma^{new}$ with probability min$\left\{1, \frac{ \phi(\bgamma \mid \mu(\bgamma^{new}), \tau^2_\gamma I_p) \phi(\bgamma^{new} \mid 0, \sigma^2_\gamma \Lambda_\gamma)\prod_i f(y_i \mid \bgamma^{new}, \bullet) }{\phi(\bgamma^{new} \mid \mu(\bgamma), \tau^2_\gamma I_p) \phi(\bgamma \mid 0, \sigma^2_\gamma \Lambda_\gamma)\prod_i f(y_i \mid \bgamma, \bullet) }\right\}$.
\item Update $\xi$: \quad Given $\bgamma$ and $\lambda$, update $\xi = \{ j: \gamma_j > \lambda\}$.

\item Update $\bfalpha$ (MALA): \quad For $j \notin \xi$, sample $\alpha_j \sim\mN(0, \sigma_\alpha^2)$. Draw $\bfalpha_\xi^{new} \sim\mN\left\{ \mu(\bfalpha_\xi), \tau^2_\alpha I_{|\xi|}\right\}$, where $\mu(\bfalpha_\xi) = \bfalpha_\xi + \frac{\tau^2_\alpha}{2} ( \frac{\partial \text{log}f}{\partial \bfalpha_\xi}- \frac{1}{2} \Sigma_\xi \bfalpha_\xi)$ with $\frac{\partial \text{log}f}{\partial \alpha_j} = \sum_{i=1}^n (a' (\bfz^{\rT}_i \bfomega + \bfx^{\rT}_i \bbeta) + b' (\bfz^{\rT}_i \bfomega + \bfx^{\rT}_i \bbeta)) x_{ij}$ for $j \in \xi$ and $\Sigma_\xi = \sigma_\alpha^2 I_{|\xi|}$. Update $\bfalpha_\xi  \longleftarrow \bfalpha_\xi^{new}$ with probability

min$\left\{1, \frac{ \phi(\bfalpha_\xi \mid \mu(\bfalpha_\xi^{new}), \tau^2_\alpha I_{|\xi|}) \phi(\bfalpha_\xi^{new} \mid 0, \Sigma_\xi)\prod_i f(y_i \mid \bfalpha_\xi^{new}, \bullet) }{\phi(\bfalpha_\xi^{new} \mid \mu(\bfalpha_\xi), \tau^2_\alpha I_{|\xi|}) \phi(\bfalpha_\xi \mid 0, \Sigma_\xi)\prod_i f(y_i \mid \bfalpha_\xi, \bullet) }\right\}$.

\item  Update $\sigma_\gamma^2$: \quad Draw $\sigma_\gamma^2 \sim\mIG(a^\gamma, b^\gamma)$ where $a^\gamma = a_\gamma + \frac{p}{2}$ and $b^\gamma = b_\gamma + \frac{\bgamma^{\rT} \Lambda^{-1}_\gamma \bgamma}{2}$.
\item  Update $\sigma_\alpha^2$: \quad Draw $\sigma_\alpha^2 \sim\mIG(a^\alpha, b^\alpha)$ where $a^\alpha = a_\alpha + \frac{p}{2}$ and $b^\alpha = b_\alpha + \frac{\sum_j \alpha^2_j}{2}$.
\item Update $\varepsilon$ (Random Walk, optional) \quad Draw $\varepsilon^{new} \sim N(\varepsilon, \tau^2_\varepsilon)$. Update $\varepsilon  \longleftarrow \varepsilon^{new}$ with probability
min$\left\{1, \frac{|\bfL + \varepsilon^{new} \bfI_{p_n}|^{\frac{1}{2}} \frac{1}{\varepsilon^{new}} \exp(-\frac{\varepsilon^{new} \bgamma^T \bgamma}{2\sigma_\gamma^2} - \frac{(log\varepsilon^{new} - \mu_\varepsilon)^2}{2\sigma^2_\varepsilon}) }{|\bfL + \varepsilon \bfI_{p_n}|^{\frac{1}{2}} \frac{1}{\varepsilon} \exp(-\frac{\varepsilon \bgamma^T \bgamma}{2\sigma_\gamma^2} - \frac{(log\varepsilon - \mu_\varepsilon)^2}{2\sigma^2_\varepsilon})}\right\}$.
\item  Update $\lambda$: \quad Given $\lambda$, draw $\lambda^{new} \sim N_{+} (\lambda, \lambda_l, \lambda_u, \sigma^2_l)$. Set $\lambda \longleftarrow \lambda ^{new}$ with probability min \\
$\left\{1, \frac{ \phi_{+}(\lambda \mid \lambda^{new}, \lambda_l, \lambda_u, \sigma^2_l) \prod_i f(y_i \mid \lambda^{new},\bullet) }{\phi_{+}(\lambda^{new} \mid \lambda, \lambda_l, \lambda_u, \sigma^2_l) \prod_i f(y_i \mid \lambda, \bullet) }\right\}$.
\end{itemize}

 The proposal variances $\tau^2_\gamma$, $\tau^2_\alpha$ and $\tau^2_\omega$ are all adaptively chosen by tuning acceptance rates to 30\% for random walk and 50\% for MALA in simulation studies and 15\% for random walk and 30\% for MALA in real data analysis. Our choice of the acceptance rate takes into account the general theoretical results on the optimal scaling of random walk~\citep{roberts1997weak} and MALA~\citep{roberts1998optimal, roberts2001optimal}. However, the log-likelihood of our model involves both smooth and discontinuous functions, which do not satisfy the regularity conditions of the general theoretical results. Thus, we have made slight changes in the theoretical optimal acceptance rates according to our numerical experiments.

Denote by $\{\bgamma^{(i)}, \bfalpha^{(i)}, \lambda^{(i)}\}_{i=1}^N$ the MCMC samples obtained after burn-in. We estimate the posterior inclusion probability for node $j (j =1 , \cdots, p_n)$ by
$$\widehat\mPr(\beta_j \neq 0 \mid D_n)= \frac{1}{N} \sum_{i=1}^N I\{|\gamma_j^{(i)}|>\lambda^{(i)}\}.$$
According to \cite{barbieri2004optimal}, we select the informative nodes with at least $50\%$ inclusion probability, denote by $\widehat M = \{j: \widehat\mPr(\beta_j \neq 0 \mid D_n) > 0.5\}$ the indices of all the informative nodes.  To estimate regression coefficients of informative nodes, we choose the estimated conditional expectation of $\beta_j$ given $\beta_j \neq 0$ by
$$
\widehat\mE\{\beta_j\mid \beta_j \neq 0, D_n\}= \frac{\sum_{i=1}^N \alpha_j^{(i)} I (|\gamma_j^{(i)}|>\lambda^{(i)})}{\sum_{i=1}^N I\{|\gamma_j^{(i)}|>\lambda^{(i)}\}}, \mbox{for } j \in \widehat M.
$$

\section{Numerical Studies}
\label{sec5}
We conduct simulation studies to evaluate performance of the proposed methods compared with existing methods for different scenarios.

\subsection{Small simple networks}

\begin{figure}[htbp]
  \centering
\begin{tabular}{ccc}
\includegraphics[width=1.5in]{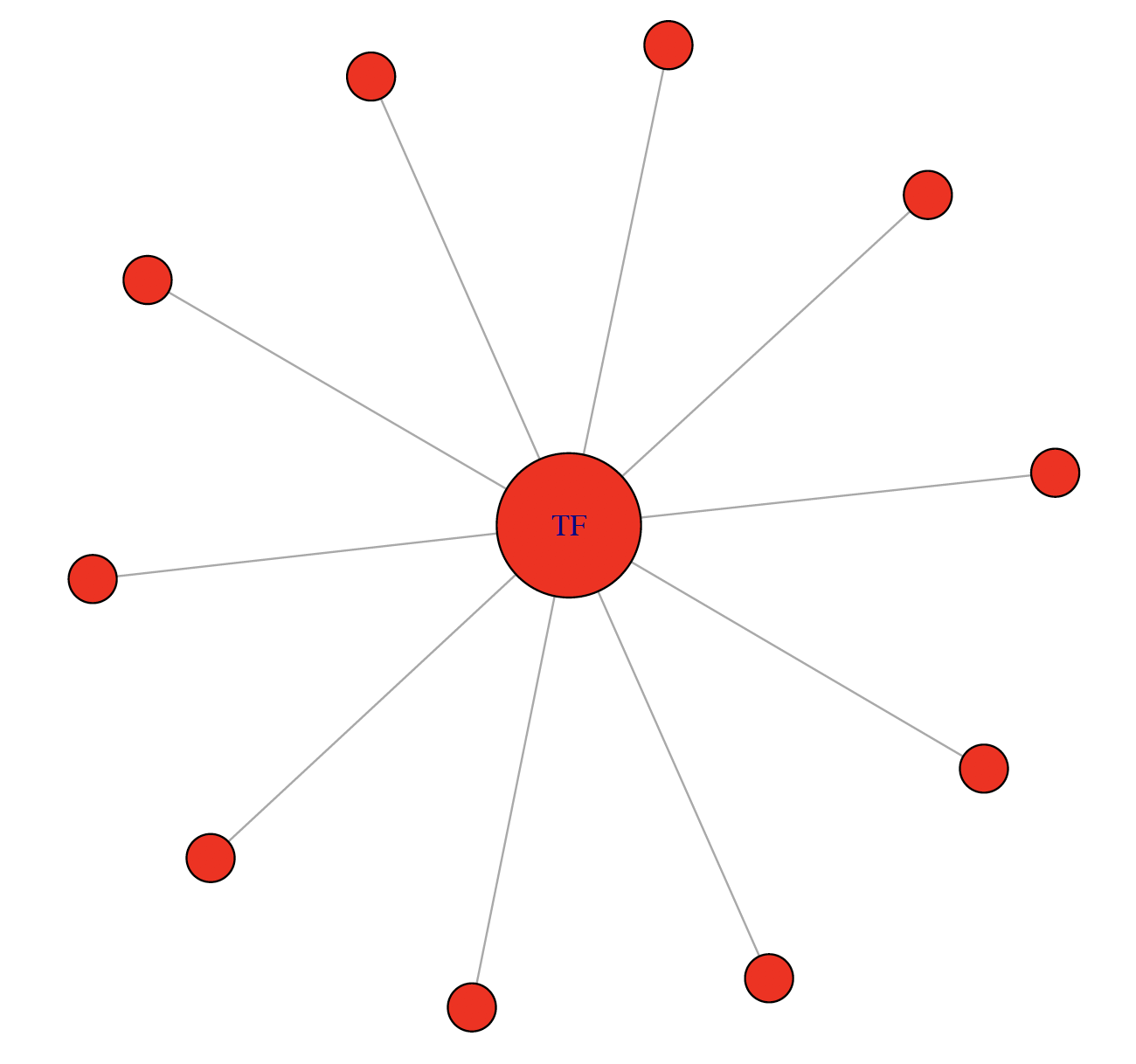} &
\includegraphics[width=1.5in]{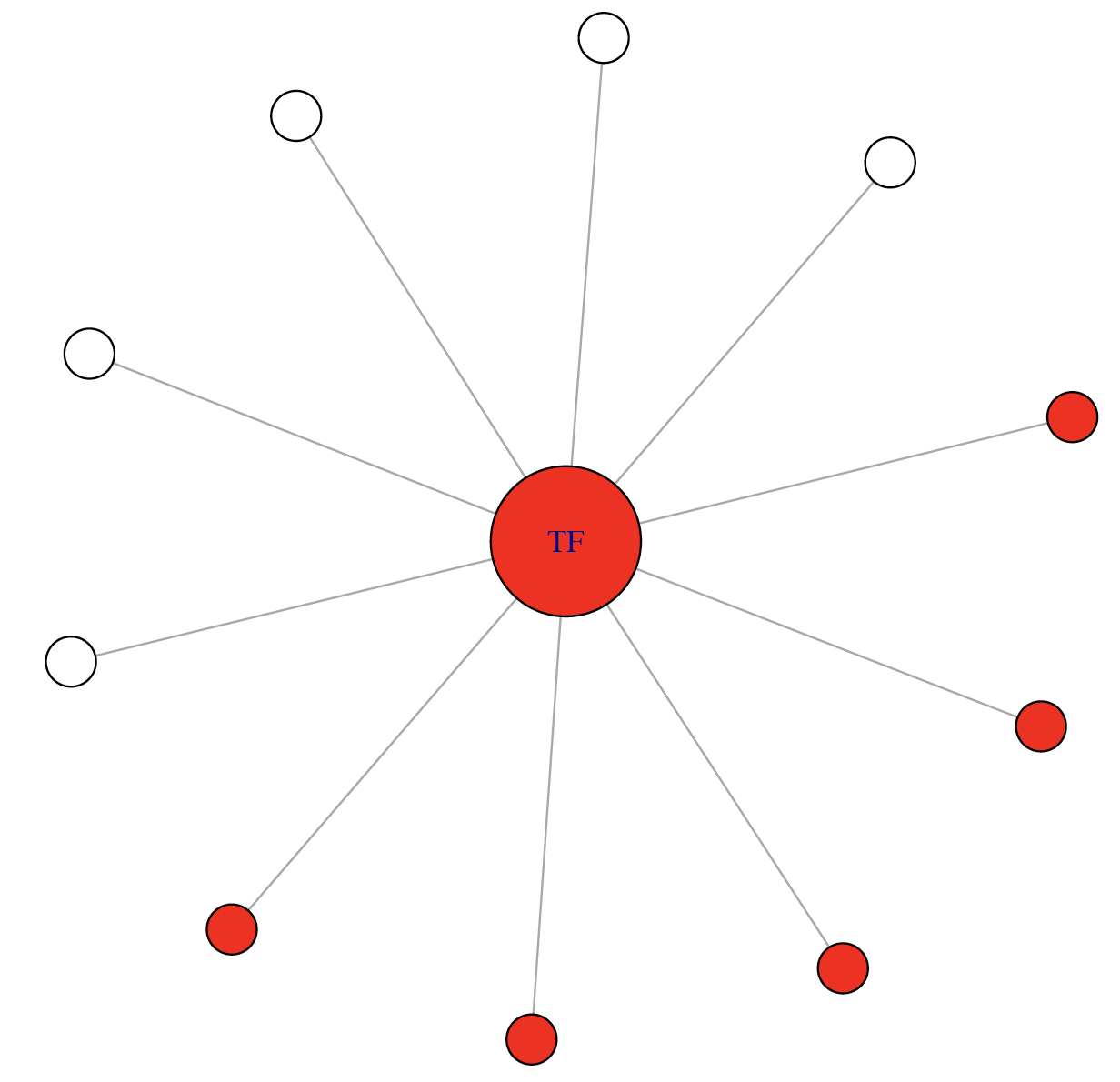} \\
(a) Type 1& (b) Type 2
\end{tabular}
\caption{Two types network markers in the simulated small simple networks, where true informative nodes are marked in red. In Type 1 network marker, TF and all target genes are informative nodes. In Type 2 network marker, TF and half of target genes are informative nodes.}
\label{case1} %% label for entire figure
\end{figure}

Following settings in  \cite{li2008network},  \cite{zhe2013joint} and \cite{kim2013network}, we simulate small simple gene networks consisting of multiple subnetworks, where each subnetwork contains one transcription factor (TF) gene and 10 target genes that are connected to the TF gene; and two of the subnetworks are set as the true network markers.  We consider two types of true network markers. In Type 1 network marker, TF and all 10 target genes are informative nodes; see Figure \ref{case1}(a).  In Type 2 network marker, TF and half of target genes are informative nodes; see Figure \ref{case1}(b). For each informative node, the magnitude of the effect size, $\beta$, is simulated from $\text{Unif}(1,3)$ and its sign is randomly assigned as positive or negative.

In each subnetwork, the covariate variables for 11 nodes, i.e., the expression levels of the TF gene and 10 target gene, are jointly generated from a 11-dimensional multivariate normal distribution with zero mean and unit variance, where the correlation between the TF gene and each target gene is $0.5$; and the correlation between any two different target genes is 0.25. We assume the covariate variables are independent across different subnetworks.

%The  TF gene expression levels, denoted $X_{\mathrm{TF}}$, is generated from the standard normal distribution.  Given the $X_{\mathrm{TF}}$, each target gene, denoted $X_{\mathrm{tg}}$, is independently sampled from  a normal distribution with mean $0.5 X_{\mathrm{TF}}$ and variance $0.75$. This implies that $X_{\mathrm{TF}}$ and multiple $X_{\mathrm{tg}}$'s jointly follow a multivariate normal distribution with zero mean and unit variance, where the correlation between $X_{\mathrm{TF}}$ and $X_{\mathrm{tg}}$ is $0.5$; and the correlation between any two different $X_{\mathrm{tg}}$'s is 0.25.

 To generate the response variable given the true network markers, we consider binary and continuous cases, where the continuous response variable is generated from linear regression, i.e. $y \sim\mN(X \bbeta, \sum_i \beta_i^2/3)$; and the binary response is generated from logistic regression, i.e. $\mPr(y = 1) = 1/\{1+\exp(-X\bbeta)\}$.

We consider two scenarios for the number of subnetworks: 3 and 10; the corresponding numbers of nodes, $p = 33$ and $p = 110$ respectively.  For the network with Type 1 markers, the number of informative nodes is 22; For the network with Type 2 markers, the number of informative nodes is 12.  We generate 50 datasets for each scenario. For linear regression, each dataset contains 100 training samples and 100 test samples; for logistic regression,  each dataset contains 200 training samples and 200 test samples.

We compare the proposed TGLG approach with the following existing methods: Lasso \citep{tibshirani1996regression}, Elastic-net \citep{zou2005regularization}, Grace \citep{li2008network}, aGrace \citep{li2010variable}, $L_\infty$ and a$L_\infty$  \citep{luo2012two}, TTLP and LTLP \citep{kim2013network}, BRGL \citep{liu2014bayesian} and Ising model \citep{goldsmith2014smooth, li2015spatial}. For the hyper priors in the TGLG approach, we assign weakly informative priors: $\sigma^2_\gamma \sim \mIG(0.01, 0.01), \sigma^2_\alpha \sim\mIG(0.01, 0.01)$. For all the regularized approaches, we adopt three-fold cross validations to choose tuning parameters. For the Ising prior model, we specify the priors as $$p(\bfgamma) = \phi(a, b) \exp \left[a\sum_i \gamma_i + \sum_i \left\{ \sum_{j \in N_i} b I(\gamma_i = \gamma_j)\right\}\right ]$$ and $\beta_i | \gamma_i =1 \sim \mN(0, \sigma^2_\beta)$, where $N_i$ denotes the neighbor nodes set of node $i$. For hyper prior specifications in Ising model, we fix $a=-2$ and choose $b$ from $2, 5, 7$ and $10$ based on model performance. We implement a single-site Gibbs sampler for Ising model. For BRGL by~\cite{liu2014bayesian}, the network markers are selected when the posterior probability $\mP(|\beta_j|>\sqrt{\mathrm{Var}(\beta_j)} | D_n)$ exceeds 0.5.

To evaluate posterior sensitivity to the prior specification of $\bfgamma$ in TGLG, we consider three cases. 1) TGLG-I: assign a network-independent prior for $\bfgamma$, i.e., $\bfgamma \sim \mN\{\bfzero_{p}, \sigma_\gamma^2I_p)$; 2) TGLG-F: fix $\varepsilon = 10^{-5}$ and 3) TGLG-L: assign a log-normal priors to  $\varepsilon$, i.e., $\log\varepsilon \sim N(-5,9)$.

For all the Bayesian methods, we run $30,000$ MCMC iterations with the first $20,000$ as burn-in. We also check the MCMC convergences by running five chains and computing the Gelman-Rubin diagnostics. For each of the Bayesian methods, the estimated 95\% CI of the potential scaled reduction factors is $[1.0, 1.0]$, indicating the convergence of the MCMC algorithm.  To compare the performance of different methods, we compute true positives, false positives and the area under the curve (AUC) for true network markers recovery, prediction mean squared error (PMSE) for linear regression and classification  error (CE) for logistic regression regarding to outcome. We report the mean and standard error over 50 datasets for each metric we choose to compare in the result table.

Table \ref{linear} summarizes the results for linear regression under different settings.  In most cases,  TGLG approaches with incorporating network structure achieve a smaller PMSE,  smaller number of false positives with a comparable amount of true positives  compared with other methods. For the Ising model, we only report the results in the case of $b=7$ since it has an overall best performance among all choices of $b$ values. In fact, the performance of the Ising model varies greatly for different choices of values for $b$ and it may perform vary bad with an inappropriate value of $b$. Table \ref{time} shows the mean computation time over 50 datasets for Ising model and TGLG. It shows that our method is much more computationally efficient than the Ising model, especially for the large-scale networks.

As for the three cases of adopting TGLG approaches, TGLG-L has the best overall performance regarding to the PMSE and false positives. TGLG-F tends to have a larger false positive than TGLG-L and TGLG-I, since selection variables for connected nodes are highly dependent when fixed $\varepsilon = 10^{-5}$. However, TGLG-F still has a smaller PMSE than TGLG-I. Compared with TGLG-I, TGLG-L has smaller false positives and PMSE in most cases. These facts show that incorporating network structure can improve model prediction performance of TGLG in linear regression.

Table \ref{logic} summarizes the results for the logistic regression under different simulation settings. Here the TGLG is only compared with Lasso, Elastic-net and the Ising model. For Type 1 network, the Ising model has a smaller number of false positives  than all three TGLG approaches. However, the Ising model has a larger prediction error and a smaller  number of true positives. In all other scenarios, TGLG outperforms the Ising model. Table \ref{time} demonstrates the TGLG approach is much more computational efficient than the Ising model in Logistic regression. In addition, TGLG-F and TGLG-L have a smaller number of false positives and classification error than TGLG-I in most cases, which indicates that including network structure could improve model performance in logistic regression.

\renewcommand{\arraystretch}{1.1}
\begin{table}
\caption{\label{linear} Simulation results for linear regression. PMSE: prediction mean squared error. TP: true positives, FP: false positives; number of informative nodes in Type 1 network is 22; number of informative nodes in Type 2 network is 12.}
  \tiny
  \centering
\begin{tabular}{llllllllll}
\hline
  Method & PMSE & TP & FP &AUC &  & PMSE & TP & FP& AUC\\
\hline
  &\multicolumn{4}{c}{Type 1 $p=33$} & & \multicolumn{4}{c}{Type 1 $p=110$}\\
 \hline
   Lasso& 52.3(1.6)&20.6(0.2)&7.3(0.3)  &0.778(0.006) & & 71.6(1.9)&17.2(0.3)&	19.6(1.2)& 0.792(0.005) \\
  Elastic-net& 50.9(1.4)&21.8(0.1)&10.4(0.2)  &0.788(0.004) & & 73.7(1.8)&19.6(0.3)&	46.6(2.9)& 0.811(0.004)\\
  Grace& 56.8(1.5)&21.6(0.1)&10.1(0.2)  &0.864(0.007) & & 87.5(2.0)&17.9(0.4)&	37.5(2.5)& 0.897(0.004)\\
    aGrace& 53.7(1.5)&22.0(0.0)&10.7(0.1)  &0.875(0.007) & & 76.4(2.1)&20.6(0.3)&	65.9(3.6)&0.899(0.005) \\
  $L_\infty$& 51.4(1.5)&21.8(0.1)&8.9(0.4)  & 0.970(0.006)& & 66.5(1.7)&21.5(0.2)&	22.7(1.5)&0.973(0.005) \\
  a$L_\infty$& 54.2(1.3)&21.8(0.1)&8.2(0.6)  & 0.669(0.034)& & 63.5(1.5)&21.5(0.2)&	19.6(1.4)& 0.946(0.010)\\
  TTLP& 54.3(1.6)&21.9(0.0)&10.1(0.4)  &0.834(0.019) & & 72.6(2.0)&20.9(0.4)&	44.2(4.6)& 0.920(0.004)\\
  LTLP& 51.3(1.2)&22.0(0.0)&8.8(0.6)  & 0.933(0.005)& & 67.1(1.7)&21.5(0.2)&	57.6(2.7)&0.897(0.009) \\
BRGL & 51.0(1.3)&19.5(0.2)&4.1(0.3)  & 0.883(0.008) & & 79.7(1.8)&17.9(0.2)&	22.1(0.9) & 0.867(0.006) \\
  Ising(b=7) & 54.9(3.0)&19.7(0.7)&2.9(0.7)  & 0.925(0.017) & & 94.9(5.9)&15.1(0.9)&	33.9(2.4) & 0.786(0.023) \\
TGLG-I & 50.1(1.3)&21.9(0.1)&10.7(0.2)  & 0.863(0.010) & &81.4(2.1)&14.8(0.5)&	22.6(2.6)& 0.779(0.009) \\
 TGLG-F & 45.2(1.2)&22.0(0.0)&2.2(0.6)  &0.912(0.032) & & 63.9(2.8)&19.7(0.4)&	17.8(2.9) &0.899(0.016) \\
 TGLG-L& 46.0(1.3)&21.9(0.1)&1.7(0.5)  & 0.968(0.016) & & 74.1(2.4)&17.1(0.5)&	19.3(2.7) & 0.847(0.013) \\

  \hline
 &\multicolumn{4}{c}{Type 2 $p=33$} & & \multicolumn{4}{c}{Type 2 $p=110$}\\
 \hline
  Lasso& 23.1(0.6)&11.7(0.1)&11.8(0.6)  &0.830(0.006)& & 30.6(0.8)&9.5(0.2)&	19.1(1.1)& 0.826(0.007)\\
  Elastic-net& 23.4(0.6)&11.8(0.1)&15.4(0.6)  &0.802(0.006) && 31.4(0.9)&10.6(0.2)&	34.0(2.1)& 0.818(0.006)\\
  Grace& 25.8(0.6)&11.4(0.1)&14.7(0.6)  &0.813(0.005) & & 35.2(0.8)&9.1(0.2)&	25.8(1.9)& 0.855(0.005)\\
    aGrace& 25.9(0.7)&12.0(0.0)&20.3(0.3)  &0.868(0.006) & & 32.8(0.8)&11.6(0.1)&	73.0(3.5)& 0.895(0.007)\\
  $L_\infty$& 23.8(0.6)&11.9(0.1)&17.2(0.6)  &0.812(0.005) & & 30.3(0.7)&11.3(0.2)&	28.9(1.9)& 0.928(0.005)\\
  a$L_\infty$& 26.1(0.7)&11.9(0.1)&16.9(0.6)  &0.643(0.018) & & 30.6(0.6)&11.3(0.2)&	27.1(1.7)& 0.893(0.009)\\
  TTLP& 25.9(0.8)&12.0(0.0)&20.0(0.5)  &0.801(0.008) & & 32.2(0.8)&11.6(0.2)&	64.3(5.2)& 0.923(0.004)\\
  LTLP& 24.7(0.7)&12.0(0.0)&20.4(0.4)  &0.825(0.008) & & 30.6(0.7)&11.7(0.2)&	75.1(3.6)& 0.864(0.006)\\
BRGL  & 23.7(0.6)&11.4(0.1)&7.3(0.4)&0.938(0.007) & & 37.7(0.9)&9.9(0.1)&	23.8(1.1)& 0.876(0.008) \\
  Ising(b=7) & 27.8(1.5)&9.9(0.5)&11.6(0.8)  & 0.855(0.024) & & 45.8(2.6)&7.6(0.6)&	44.5(2.0) & 0.709(0.032) \\
  TGLG-I& 23.7(0.6)&10.8(0.2)&8.0(0.9) &0.918(0.006) & & 33.9(0.9)&7.2(0.3)&	7.6(1.5)& 0.829(0.011) \\
TGLG-F& 22.8(0.6)&11.4(0.1)&10.2(0.7)&0.901(0.015)  & & 28.7(1.1)&10.5(0.3)&	14.2(2.1)& 0.922(0.012)\\
 TGLG-L & 22.3(0.6)&11.6(0.1)&8.9(0.6)&0.930(0.008) & & 28.8(0.9)&8.8(0.3)&	6.4(1.1)& 0.908(0.011) \\

\hline
\end{tabular}

\end{table}

\begin{table}
\caption{\label{logic}Simulation results for logistic regression with sample size 200. CE: classification error (number of incorrect classification); TP: true positive; FP: false positive; number of Type 1 true network markers: 22; number of Type 2 true network markers:  12.}
\centering
{\tiny
\begin{tabular}{llllllllll}
\hline
  Method & CE & TP & FP &AUC&  & CE & TP & FP&AUC\\
   \hline
  &\multicolumn{4}{c}{Type 1 $p=33$} & & \multicolumn{4}{c}{Type 1 $p=110$}\\
 \hline
  Lasso& 20.8(0.7)&21.2(0.1)&6.9(0.4)  &0.811(0.004)& & 30.8(1.1)&19.1(0.4)&	25.1(1.7)& 0.836(0.004)\\
  Elastic-net& 21.0(0.8)&21.4(0.1)&8.4(0.4)  &0.818(0.004)& & 32.6(0.8)&19.9(0.2)&	29.4(2.1)& 0.848(0.003)\\
 Ising(b=5) & 39.2(3.0)&15.2(1.2)&0.0(0.0)  & 0.937(0.011) & & 47.6(4.1)&13.5(1.1)&	10.2(2.9) & 0.826(0.031) \\
 TGLG-I& 19.2(0.6)&21.9(0.1)&10.0(0.2) &0.877(0.011) & & 30.5(0.9)&17.1(0.3)&	16.0(1.4) &0.851(0.008)\\
 TGLG-F& 19.4(0.7)&21.8(0.1)&8.0(0.5)&0.858(0.021)  & & 30.8(1.1)&17.6(0.4)&	13.0(1.1) &0.870(0.007)\\
 TGLG-L& 18.7(0.7)&21.8(0.1)&7.5(0.5) &0.875(0.018) & & 30.4(1.0)&17.3(0.3)&	13.4(1.1)&0.858(0.008) \\

\hline
 &\multicolumn{4}{c}{Type 2 $p=33$} & & \multicolumn{4}{c}{Type 2 $p=110$}\\
  \hline
   Lasso& 25.2(0.9)&11.7(0.1)&10.1(0.7) & 0.856(0.004)& & 32.7(1.0)&10.6(0.2)&	22.7(2.2) &0.872(0.004)\\
  Elastic-net& 26.1(0.8)&11.9(0.0)&13.2(0.7) &0.796(0.004) & & 36.6(1.2)&10.5(0.3)&	25.9(2.5)& 0.849(0.004)\\
     Ising(b=5) & 27.4(1.4)&9.5(0.4)&7.2(0.4)  & 0.899(0.016) & & 37.7(2.8)&7.4(0.5)&	9.0(1.7) & 0.820(0.025) \\
	TGLG-I& 22.6(0.8)&11.4(0.1)&4.8(0.6) &0.961(0.007)  & & 29.4(1.2)&9.7(0.3)&	6.9(0.9)&0.897(0.0012) \\
 TGLG-F & 23.2(0.8)&11.5(0.1)&6.3(0.6)&0.941(0.010)  & & 29.3(1.0)&9.9(0.3)&	6.7(0.6) &0.903(0.010)\\
 TGLG-L & 22.1(0.8)&11.6(0.1)&5.8(0.7)&0.959(0.005)  & & 28.6(1.0)&10.1(0.2)&	6.2(0.8)&0.921(0.009) \\

\hline

\end{tabular}
}

\end{table}

\begin{table}

\caption{\label{time}{ {Average computing time with standard error in seconds for Ising model and TGLG based network marker selection. All the computations run on a desktop computer with $3.40$ GHz i7 CPU and $16$ GB memory}}}
\centering
  {\footnotesize
\begin{tabular}{llllll}
\hline
&&\multicolumn{2}{c}{Linear regression} &  \multicolumn{2}{c}{Logistic regression}\\
\hline
  &                 &Ising & TGLG& Ising & TGLG\\
\hline
$p=33$&Type 1  & 140.1(0.5) & 21.5(0.2) & 230.1(7.6) & 26.7(0.3) \\
&Type 2  & 140.1(0.5) & 21.0(0.3) & 229.9(7.6) & 26.4(0.2) \\[1mm]
$p=110$&Type 1  & 1191.4(7.1) & 31.7(0.2) & 1210.1(10.1) & 37.7(1.0) \\
&Type 2 & 1153.4(8.5) & 30.6(0.1) & 1203.6(8.4) & 36.5(0.9) \\
\hline
\end{tabular}
}
\end{table}

\subsection{Large scale-free networks}
We perform simulation studies on large scale-free networks, which are commonly used network models for gene networks.  We simulate  scale-free network~\citep{barabasi1999emergence} with 1,000 nodes using  {\tt R} function {\tt barabasi.game} in package {\tt igraph}. In the simulated scale-free network, we set the true network markers by selecting 10 nodes out of 1,000 as the true informative nodes according to two criteria: 1) all the true informative nodes form a connected component~\citep{hopcroft1973algorithm} in the network; 2) all the true informative nodes are disconnected, in which case the TGLG model assumption does not hold.  For each informative node, the magnitude of the effect size is simulated from $\text{Unif}(1,3)$ and its sign is randomly assigned as positive or negative. Covariates $X$ are generated from a multivariate normal distribution $X \sim \mN(0, 0.3^D)$, where $D$ is the shortest path distance matrix between nodes in the generated scale-free network. Response variable $Y$ is generated using $Y \sim \mN(X \bbeta, \sum \beta_i^2/3)$ for linear regression and $\mPr(Y = 1) = 1/\{1+\exp(-X\bbeta)\}$ for logistic regression. According to the above procedure, we simulate 50 datasets with sample size 200.

We apply the aforementioned all three TGLG methods (TGLG-I, TGLG-F, TGLG-L) to each dataset compared with Lasso and Elastic-net. In addition, to evaluate the robustness of network structure mis-specifications, for each simulated scale-free network, we randomly select 20\% of nodes and permute their labels; and then we apply TGLG-L with this mis-specified network. We refer to this approach as TGLG-M.

%Another concern for TGLG is that TGLG prior construction depends on the correctly specified network structure.

Table \ref{scale} reports the same performance evaluation metrics as Table \ref{linear} and Table \ref{logic}.  When the informative nodes form a connected component in the network, overall TGLG-L achieve the best performance regarding to PMSE or CE, and false positives. When the informative nodes are all disconnected, TGLG-L still has the best performance in linear regression, but is slightly worse than TGLG-I in logistic regression. This fact indicates that TGLG approaches is not sensitive to our model assumption regarding the true network markers.  In both cases,  TGLG-M performs worse than TGLG-L with correctly specified networks, but still better than Lasso and Elastic-net. This implies that the useful network information can improve the performance of TGLG, while TGLG-L is robust the network misspecification.
 \begin{table}
\caption{\label{scale}Simulation results for scale-free network. The number of true informative nodes is 10. Sample size is 200 and the number of nodes is 1,000. }
\centering

  {\tiny
\begin{tabular}{llllllllll}
\hline
  Method & PMSE & TP & FP &AUC&  & CE & TP & FP&AUC\\
   \hline
  &\multicolumn{4}{c}{Linear regression} & & \multicolumn{4}{c}{Logistic regression}\\
 \hline
 &\multicolumn{9}{c}{True informative nodes form a connected component} \\
 \hline
  Lasso& 21.7(0.6)&9.5(0.1)&54.4(3.8)& 0.936(0.004)& & 43.3(1.6)&8.4(0.2)&	29.6(3.4)& 0.771(0.028)\\
  Elastic-net& 23.2(0.7)&9.6(0.1)&69.0(3.9) & 0.931(0.004)& & 57.9(2.4)&7.7(0.2)&	22.4(3.2)& 0.928(0.006)\\
	TGLG-I& 21.7(0.8)&9.1(0.1)&13.5(1.9)&0.950(0.007)  & & 37.2(1.3)&7.7(0.2)&	8.9(0.9)&0.892(0.011) \\
 TGLG-F& 21.8(0.9)&9.3(0.1)&14.6(1.5)&0.968(0.006)  & & 35.2(1.3)&8.0(0.2)&	7.8(0.9)&0.902(0.011) \\
 TGLG-L& 20.7(0.7)&9.1(0.1)&10.1(1.5)&0.957(0.006)  & & 35.4(1.4)&7.9(0.3)&8.3(1.0)&0.893(0.011) \\
 TGLG-M & 21.2(0.8)&9.1(0.1)&11.3(1.5)&0.952(0.007)  & & 37.1(1.3)&7.8(0.2)&9.3(1.1)&0.892(0.012) \\
\hline
   &\multicolumn{9}{c}{True informative nodes are all disconnected} \\
 \hline
  Lasso& 20.8(0.6)&9.8(0.1)&55.0(3.7)& 0.940(0.003) & & 43.4(1.2)&8.9(0.2)&	26.8(3.0)& 0.824(0.028)\\
  Elastic-net& 22.2(0.7)&9.8(0.1)&68.6(3.9) & 0.941(0.003)& & 55.7(1.9)&8.4(0.2)&	27.3(4.0)& 0.939(0.003)\\
	TGLG-I & 21.4(0.9)&9.4(0.1)&13.4(2.0)&0.974(0.006)  & & 35.4(1.3)&8.6(0.2)&	7.9(0.8)&0.931(0.009) \\
 TGLG-F& 21.7(0.8)&9.4(0.1)&16.7(1.9)&0.971(0.006)  & & 35.5(1.4)&8.4(0.2)&	7.8(0.9)&0.922(0.010) \\
 TGLG-L & 20.6(0.8)&9.6(0.1)&11.6(2.1)&0.980(0.004)  & & 36.9(1.5)&8.5(0.2)&	9.4(1.1)&0.925(0.009) \\
  TGLG-M & 21.3(0.9)&9.4(0.1)&11.4(1.7)&0.969(0.005)  & & 35.3(1.2)&8.5(0.2)&	8.4(0.9)&0.928(0.008) \\
\hline
\end{tabular}
}

\end{table}

\subsection{Application to breast cancer data from the Cancer Genome Atlas}
In the real data application, we use the High-quality INTeractomes (HINT) database for the biological network \citep{Das:2012aa}. We apply our method to the TCGA breast cancer (BRCA) RNA-seq gene expression dataset with $762$ subjects and $10,792$ genes in the network. The response variable we consider here is ER status - whether the cancer cells grow in response to the estrogen. The ER status is a molecular characteristic of the cancer which has important implications in prognosis. The purpose here is not focused on prediction. Rather we intend to find genes and functional modules that are associated with ER status, through which biological mechanisms differentiating the two subgroups of cancer can be further elucidated.

We code ER-positive as 1 and ER-negative as 0. We remove subjects with unknown ER status. In total, there are 707 subjects with 544 ER-positive and 163 ER-negative. We remove 348 gene nodes with low count number, which leaves us with 10,444 nodes. To apply our methods, we first standardize the  gene nodes and then apply a logistic regression model for network marker selection. For prior settings, we use $\sigma^2_\gamma \sim\mIG(0.01, 0.01), \sigma^2_\alpha \sim\mIG(0.01, 0.01)$ and $\sigma^2_\omega = 50$. We fix $\lambda$ at different grid values and choose $\lambda = 0.004$ by maximizing the likelihood values. The MCMC algorithms runs 100,000 iterations with first 90,000 as burn-in and thin by 10. We run the chain with 3 different initial values and the Gelman-Rubin diagnostic statistic is [1.07,1.15], which shows convergence of the chain.

 A total of 470 genes are selected as networks marker by our approach. To facilitate data interpretation, we conduct the community detection on the network containing the selected network markers and their one-step neighbors \citep{clauset2004finding}. There is a total of eight modules that contain 10 or more selected genes. The plot of the modules, together with their over-represented biological process as identified using the `GOstats' package \citep{Falcon:2007aa}, are listed in Supplementary File 2.

 Figure \ref{f_results} shows two example network modules. The first example (Figure \ref{f_results}(a)) contains 95 selected gene network markers, including 14 that are connected with other network markers.  The top 5 biological processes associated with these 95 genes are listed in Table \ref{goterm}. The most significant biological process that is over-represented by the selected genes in this module is regulation of cellular response to stress (p=0.00016), with 14 of the selected genes involved in this biological process. Besides the general connection between stress response and breast cancer, ER status has some specific interplay with various stress response processes. For example, breast cancer cells up-regulate hypoxia-inducible factors, which cause higher risk of metastasis \citep{pmid24156323}. Hypoxia inducible factors can influence the expression of estrogen receptor  \citep{pmid27823907}. In addition, estrogen changes the DNA damage response by regulating proteins including ATM, ATR, CHK1, BRCA1, and p53 \citep{pmid24860786}. Thus it is expected that DNA damage response is closely related to ER status.

 Five other genes in this module are involved in the pathway of regulation of anion transport, which include the famous mTOR gene, which is implicated in multiple cancers \citep{Le-Rhun:2017aa}. The PI3K/AKT/mTOR pathway is an anticancer target  in ER+ breast cancer \citep{Ciruelos-Gil:2014aa}. The other four genes, ABCB1 \citep{Jin:2017aa}, SNCA \citep{Li:2018aa}, IRS2 \citep{Yin:2017aa} and  NCOR1 \citep{Lopez:2016aa} are all involved in some other types of cancer.

 In ER- breast cancer cells, the lack of ER signaling triggers the epigenetic silencing of downstream targets \citep{Leu:2004aa}, which explains the significance of the biological process "negative regulation of gene silencing". Many genes in the "cardiac muscle cell development" processes are also part of the growth factor receptor pathway, which has a close interplay with estrogen signaling \citep{Osborne:2005aa}. Four of the genes fall into the process "regulation of B cell proliferation". Among them, AHR has been identified as a potential tumor suppressor \citep{Formosa:2017aa}. ER$\alpha$ is recruited in AhR signaling \citep{Matthews:2006aa}. IRS2 responds to interleukin 4 treatment, and its polymorphism is associated with colorectal cancer risk \citep{Yin:2017aa}. CLCF1 signal tranduction was found to play a critical role in the growth of malignant plasma cells \citep{Burger:2003aa}. It appears that these genes are found due to their functionality in signal transduction, rather than specific functions in B cell proliferation.

\begin{figure}[htbp]
  \centering
\subfigure[]{
\label{fig:subfig:a} %% label for first subfigure
\includegraphics[width=2.5in, angle=0]{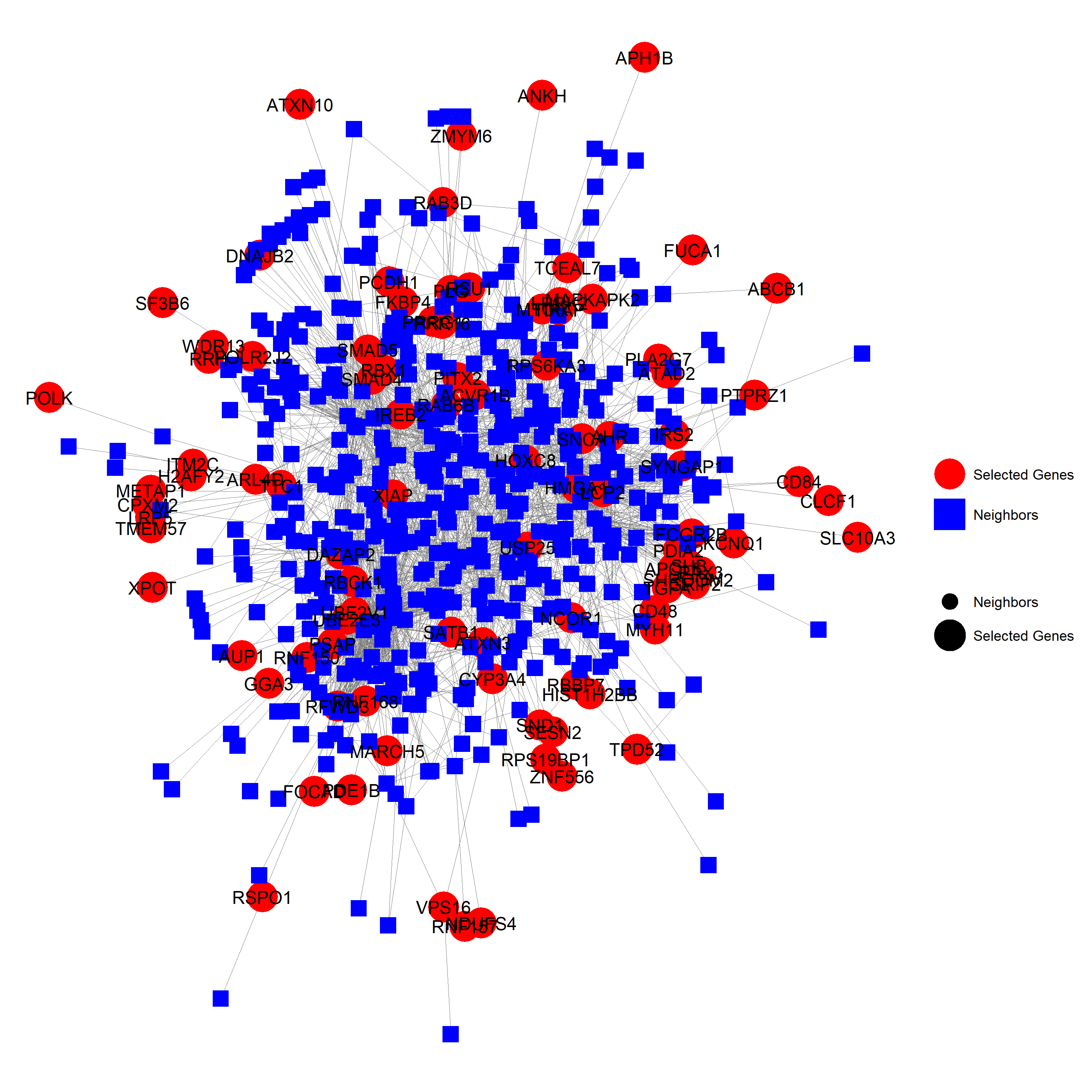}}
\hspace{-0.15in}
\subfigure[]{
\label{fig:subfig:b} %% label for second subfigure
\includegraphics[width=2.5in, angle=0]{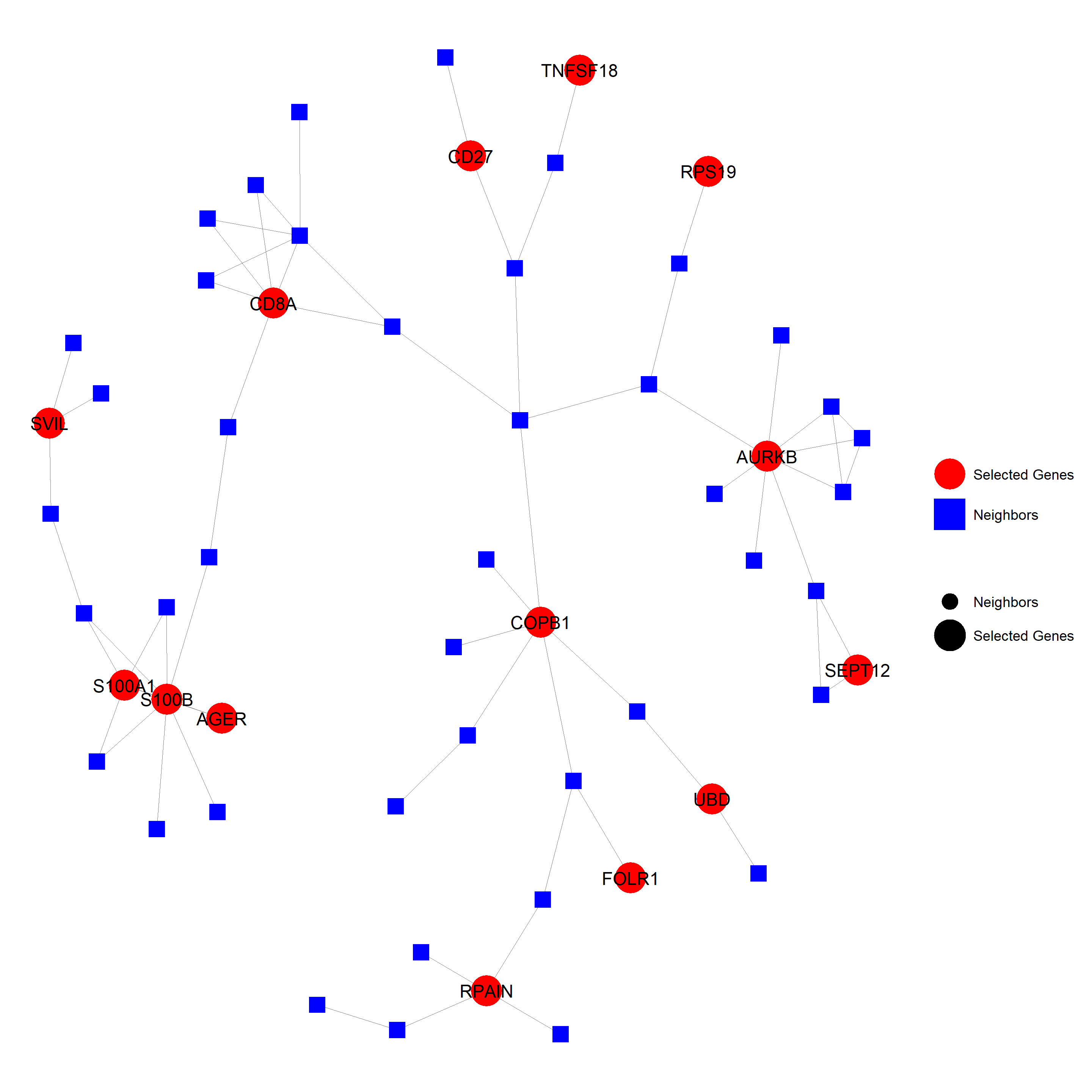}}
\hspace{-0.15in}

  \caption{\label{f_results} {\footnotesize{Two example modules of selected genes.}}}
%% label for entire figure
\end{figure}

\begin{table}
\caption{\label{goterm} Selected goterm results for the two selected modules shown in Figure \ref{f_results}. The upper part is the Goterm results for Figure \ref{f_results}(a) and the lower part is the Goterm results for Figure \ref{f_results}(b).}
\centering
{\small
\begin{tabular}{lll}
\hline
    GOBPID & Pvalue & Term\\
   \hline
  GO:0080135& 0.0001618 &regulation of cellular response to stress \\
  GO:0044070&0.000381 & regulation of anion transport \\
 GO:0060969&0.0004409 & negative regulation of gene silencing \\
 GO:0055013&0.000757 & cardiac muscle cell development \\
  GO:0030888&0.0009629 & regulation of B cell proliferation \\

\hline
 \hline
  GOBPID & Pvalue & Term\\
   \hline
  GO:0030097 & 0.00006398 &hemopoiesis  \\
  GO:1902533 &0.0003036 & positive regulation of intracellular signal transduction\\
 GO:0002250&0.0004063 & adaptive immune response \\
 GO:0032467  &0.0004452 & positive regulation of cytokinesis \\
  GO:0070229 &0.0005767 & negative regulation of lymphocyte apoptotic process \\
\hline
\end{tabular}
}

\end{table}
 The second example is a much smaller module including 14 selected genes. Six of the 14 genes are involved in both hemopoiesis and immune system development (Table \ref{goterm}). They are all signal transducers. Among them, AGER is a member of the immunoglobulin superfamily of cell surface receptors, which also acts as a tumor suppressor \citep{Wu:2018aa}. CD27 is a tumor necrosis factor (TNF) receptor. Treatment with the estrogen E2 modulates the expression of CD27 in the bone marrow and spleen cells \citep{Stubelius:2014aa}. TNFSF18 is a cytokine that belongs to the tumor necrosis factor (TNF) ligand family. Although its relation with estrogen and breast cancer is unclear, its receptor GITR shows increased expression in tumor-positive lymph nodes from advanced breast cancer patients \citep{Krausz:2012aa}, and is targeted by some anti-cancer immunotherapy \citep{Schaer:2012aa}. UBD is a ubiquitin-like protein, which promotes tumor proliferation by  stabilizing the translation elongation factor eEF1A1 \citep{Liu:2016aa}.

Interestingly, three of the other top biological processes are also immune processes. In normal immune cells, estrogen receptors regulate innate immune signaling pathways \citep{Kovats:2015aa}. In addition, some of the selected genes in these pathways have been found to associate with cancer. Examples include AURKB, which belongs to the family of serine/threonine kinases, and contributes to chemo-resistance and poor prognosis in breast cancer \citep{Zhang:2015aa}, and SVIL, which mediates the suppression of p53 protein and enhances cell survival \citep{Fang:2013aa}.

Overall, genes selected by TGLG are easy to interpret. Many known links exist between these genes and ER status, or breast cancer in general. Still many of the selected genes are not reported so far to be linked to ER status or breast cancer. Our results indicate they may play important roles.

\section{Discussion}
In summary, we propose a new prior model: TGLG prior for Bayesian network marker selection over large-scale networks. We show the proposed prior model enjoys large prior support for network marker selection over large-scale networks, leading to the posterior consistency. We also develop an efficient Metropolis-adjusted Langevin algorithm (MALA) for posterior computation.  The simulation studies show that our method performs better than existing regularized regression approaches with regard to the selection and prediction accuracy. Also, the analysis of TCGA breast cancer data indicates that our method can provide biologically meaningful results.

This paper leads to some obvious future work. First, we can apply the TGLG prior for network marker selection under other modeling framework such as the survival model and the generalized mixed effects model.   Second, the current posterior computation can be further improved by utilizing the parallel computing techniques within each iteration of the MCMC algorithm, for updating the massive latent variables simultaneously.  Third, another promising direction is to use the integrated nested laplace approximations (INLA) for Bayesian approximating computation taking advantages of the TGLG prior involving high-dimensional Gaussian latent variables.

\bibliography{ref}

\begin{thebibliography}{}

\bibitem[\protect\citeauthoryear{Aebersold and Mann}{Aebersold and
  Mann}{2003}]{aebersold2003mass}
Aebersold, R. and Mann, M. (2003).
\newblock Mass spectrometry-based proteomics.
\newblock {\em Nature} {\bf 422,} 198.

\bibitem[\protect\citeauthoryear{Barab{\'a}si and Albert}{Barab{\'a}si and
  Albert}{1999}]{barabasi1999emergence}
Barab{\'a}si, A.-L. and Albert, R. (1999).
\newblock Emergence of scaling in random networks.
\newblock {\em science} {\bf 286,} 509--512.

\bibitem[\protect\citeauthoryear{Barab{\'a}si, Gulbahce, and
  Loscalzo}{Barab{\'a}si et~al.}{2011}]{barabasi2011network}
Barab{\'a}si, A.-L., Gulbahce, N., and Loscalzo, J. (2011).
\newblock Network medicine: a network-based approach to human disease.
\newblock {\em Nature reviews genetics} {\bf 12,} 56.

\bibitem[\protect\citeauthoryear{Barbieri, Berger, et~al\mbox{.}}{Barbieri
  et~al.}{2004}]{barbieri2004optimal}
Barbieri, M.~M., Berger, J.~O., et~al. (2004).
\newblock Optimal predictive model selection.
\newblock {\em The annals of statistics} {\bf 32,} 870--897.

\bibitem[\protect\citeauthoryear{Bhattacharya, Pati, Pillai, and
  Dunson}{Bhattacharya et~al.}{2015}]{bhattacharya2015dirichlet}
Bhattacharya, A., Pati, D., Pillai, N.~S., and Dunson, D.~B. (2015).
\newblock Dirichlet--laplace priors for optimal shrinkage.
\newblock {\em Journal of the American Statistical Association} {\bf 110,}
  1479--1490.

\bibitem[\protect\citeauthoryear{Burger, Bakker, Guenther, Baum, Schmidt-Arras,
  Hideshima, Tai, Shringarpure, Catley, Senaldi, Gramatzki, and
  Anderson}{Burger et~al.}{2003}]{Burger:2003aa}
Burger, R., Bakker, F., Guenther, A., Baum, W., Schmidt-Arras, D., Hideshima,
  T., Tai, Y.-T., Shringarpure, R., Catley, L., Senaldi, G., Gramatzki, M., and
  Anderson, K.~C. (2003).
\newblock Functional significance of novel neurotrophin-1/b cell-stimulating
  factor-3 (cardiotrophin-like cytokine) for human myeloma cell growth and
  survival.
\newblock {\em Br J Haematol} {\bf 123,} 869--78.

\bibitem[\protect\citeauthoryear{Caldon}{Caldon}{2014}]{pmid24860786}
Caldon, C.~E. (2014).
\newblock Estrogen signaling and the dna damage response in hormone dependent
  breast cancers.
\newblock {\em Front Oncol} {\bf 4,} 106.

\bibitem[\protect\citeauthoryear{Chang, Kundu, and Long}{Chang
  et~al.}{2016}]{chang2016scalable}
Chang, C., Kundu, S., and Long, Q. (2016).
\newblock Scalable bayesian variable selection for structured high-dimensional
  data.
\newblock {\em arXiv preprint arXiv:1604.07264} .

\bibitem[\protect\citeauthoryear{Chekouo, Stingo, Guindani, Do,
  et~al\mbox{.}}{Chekouo et~al.}{2016}]{chekouo2016bayesian}
Chekouo, T., Stingo, F.~C., Guindani, M., Do, K.-A., et~al. (2016).
\newblock A bayesian predictive model for imaging genetics with application to
  schizophrenia.
\newblock {\em The Annals of Applied Statistics} {\bf 10,} 1547--1571.

\bibitem[\protect\citeauthoryear{Chung}{Chung}{1997}]{chung1997spectral}
Chung, F.~R. (1997).
\newblock {\em Spectral graph theory}, volume~92.
\newblock American Mathematical Soc.

\bibitem[\protect\citeauthoryear{Ciruelos~Gil}{Ciruelos~Gil}{2014}]{Ciruelos-Gil:2014aa}
Ciruelos~Gil, E.~M. (2014).
\newblock Targeting the pi3k/akt/mtor pathway in estrogen receptor-positive
  breast cancer.
\newblock {\em Cancer Treat Rev} {\bf 40,} 862--71.

\bibitem[\protect\citeauthoryear{Clauset, Newman, and Moore}{Clauset
  et~al.}{2004}]{clauset2004finding}
Clauset, A., Newman, M.~E., and Moore, C. (2004).
\newblock Finding community structure in very large networks.
\newblock {\em Physical review E} {\bf 70,} 066111.

\bibitem[\protect\citeauthoryear{Das and Yu}{Das and Yu}{2012}]{Das:2012aa}
Das, J. and Yu, H. (2012).
\newblock Hint: High-quality protein interactomes and their applications in
  understanding human disease.
\newblock {\em BMC Syst Biol} {\bf 6,} 92.

\bibitem[\protect\citeauthoryear{Dobra}{Dobra}{2009}]{dobra2009variable}
Dobra, A. (2009).
\newblock Variable selection and dependency networks for genomewide data.
\newblock {\em Biostatistics} {\bf 10,} 621--639.

\bibitem[\protect\citeauthoryear{Doi}{Doi}{2007}]{doi2007computer}
Doi, K. (2007).
\newblock Computer-aided diagnosis in medical imaging: historical review,
  current status and future potential.
\newblock {\em Computerized medical imaging and graphics} {\bf 31,} 198--211.

\bibitem[\protect\citeauthoryear{Falcon and Gentleman}{Falcon and
  Gentleman}{2007}]{Falcon:2007aa}
Falcon, S. and Gentleman, R. (2007).
\newblock Using gostats to test gene lists for go term association.
\newblock {\em Bioinformatics} {\bf 23,} 257--8.

\bibitem[\protect\citeauthoryear{Fan and Li}{Fan and
  Li}{2001}]{fan2001variable}
Fan, J. and Li, R. (2001).
\newblock Variable selection via nonconcave penalized likelihood and its oracle
  properties.
\newblock {\em Journal of the American statistical Association} {\bf 96,}
  1348--1360.

\bibitem[\protect\citeauthoryear{Fang and Luna}{Fang and
  Luna}{2013}]{Fang:2013aa}
Fang, Z. and Luna, E.~J. (2013).
\newblock Supervillin-mediated suppression of p53 protein enhances cell
  survival.
\newblock {\em J Biol Chem} {\bf 288,} 7918--29.

\bibitem[\protect\citeauthoryear{Formosa, Borg, and Vassallo}{Formosa
  et~al.}{2017}]{Formosa:2017aa}
Formosa, R., Borg, J., and Vassallo, J. (2017).
\newblock Aryl hydrocarbon receptor (ahr) is a potential tumour suppressor in
  pituitary adenomas.
\newblock {\em Endocr Relat Cancer} {\bf 24,} 445--457.

\bibitem[\protect\citeauthoryear{George and McCulloch}{George and
  McCulloch}{1993}]{george1993variable}
George, E.~I. and McCulloch, R.~E. (1993).
\newblock Variable selection via gibbs sampling.
\newblock {\em Journal of the American Statistical Association} {\bf 88,}
  881--889.

\bibitem[\protect\citeauthoryear{Gilkes and Semenza}{Gilkes and
  Semenza}{2013}]{pmid24156323}
Gilkes, D.~M. and Semenza, G.~L. (2013).
\newblock Role of hypoxia-inducible factors in breast cancer metastasis.
\newblock {\em Future Oncol} {\bf 9,} 1623--36.

\bibitem[\protect\citeauthoryear{Goldsmith, Huang, and Crainiceanu}{Goldsmith
  et~al.}{2014}]{goldsmith2014smooth}
Goldsmith, J., Huang, L., and Crainiceanu, C.~M. (2014).
\newblock Smooth scalar-on-image regression via spatial bayesian variable
  selection.
\newblock {\em Journal of Computational and Graphical Statistics} {\bf 23,}
  46--64.

\bibitem[\protect\citeauthoryear{Greicius, Krasnow, Reiss, and Menon}{Greicius
  et~al.}{2003}]{greicius2003functional}
Greicius, M.~D., Krasnow, B., Reiss, A.~L., and Menon, V. (2003).
\newblock Functional connectivity in the resting brain: a network analysis of
  the default mode hypothesis.
\newblock {\em Proceedings of the National Academy of Sciences} {\bf 100,}
  253--258.

\bibitem[\protect\citeauthoryear{Hopcroft and Tarjan}{Hopcroft and
  Tarjan}{1973}]{hopcroft1973algorithm}
Hopcroft, J. and Tarjan, R. (1973).
\newblock Algorithm 447: efficient algorithms for graph manipulation.
\newblock {\em Communications of the ACM} {\bf 16,} 372--378.

\bibitem[\protect\citeauthoryear{Jiang}{Jiang}{2007}]{jiang2007bayesian}
Jiang, W. (2007).
\newblock Bayesian variable selection for high dimensional generalized linear
  models: convergence rates of the fitted densities.
\newblock {\em The Annals of Statistics} {\bf 35,} 1487--1511.

\bibitem[\protect\citeauthoryear{Jin and Song}{Jin and Song}{2017}]{Jin:2017aa}
Jin, S.-S. and Song, W.-J. (2017).
\newblock Association between mdr1 c3435t polymorphism and colorectal cancer
  risk: A meta-analysis.
\newblock {\em Medicine (Baltimore)} {\bf 96,} e9428.

\bibitem[\protect\citeauthoryear{Johnson and Rossell}{Johnson and
  Rossell}{2012}]{johnson2012bayesian}
Johnson, V.~E. and Rossell, D. (2012).
\newblock Bayesian model selection in high-dimensional settings.
\newblock {\em Journal of the American Statistical Association} {\bf 107,}
  649--660.

\bibitem[\protect\citeauthoryear{Kang, Reich, and Staicu}{Kang
  et~al.}{2018}]{kang2018scalar}
Kang, J., Reich, B.~J., and Staicu, A.-M. (2018).
\newblock Scalar-on-image regression via the soft-thresholded gaussian process.
\newblock {\em Biometrika} {\bf 105,} 165--184.

\bibitem[\protect\citeauthoryear{Kim, Gao, and Tan}{Kim
  et~al.}{2012}]{kim2012multi}
Kim, J., Gao, L., and Tan, K. (2012).
\newblock Multi-analyte network markers for tumor prognosis.
\newblock {\em PLoS One} {\bf 7,} e52973.

\bibitem[\protect\citeauthoryear{Kim, Pan, and Shen}{Kim
  et~al.}{2013}]{kim2013network}
Kim, S., Pan, W., and Shen, X. (2013).
\newblock Network-based penalized regression with application to genomic data.
\newblock {\em Biometrics} {\bf 69,} 582--593.

\bibitem[\protect\citeauthoryear{Kitano}{Kitano}{2002}]{kitano2002systems}
Kitano, H. (2002).
\newblock Systems biology: a brief overview.
\newblock {\em Science} {\bf 295,} 1662--1664.

\bibitem[\protect\citeauthoryear{Kovats}{Kovats}{2015}]{Kovats:2015aa}
Kovats, S. (2015).
\newblock Estrogen receptors regulate innate immune cells and signaling
  pathways.
\newblock {\em Cell Immunol} {\bf 294,} 63--9.

\bibitem[\protect\citeauthoryear{Krausz, Fischer-Fodor, Major, and
  Fetica}{Krausz et~al.}{2012}]{Krausz:2012aa}
Krausz, L.~T., Fischer-Fodor, E., Major, Z.~Z., and Fetica, B. (2012).
\newblock Gitr-expressing regulatory t-cell subsets are increased in
  tumor-positive lymph nodes from advanced breast cancer patients as compared
  to tumor-negative lymph nodes.
\newblock {\em Int J Immunopathol Pharmacol} {\bf 25,} 59--66.

\bibitem[\protect\citeauthoryear{Kundu, Shin, Cheng, Manyam, Mallick, and
  Baladandayuthapani}{Kundu et~al.}{2015}]{kundu2015bayesian}
Kundu, S., Shin, M., Cheng, Y., Manyam, G., Mallick, B.~K., and
  Baladandayuthapani, V. (2015).
\newblock Bayesian variable selection with structure learning: Applications in
  integrative genomics.
\newblock {\em arXiv preprint arXiv:1508.02803} .

\bibitem[\protect\citeauthoryear{Le~Rhun, Bertrand, Dumont, Tresch, Le~Deley,
  Mailliez, Preusser, Weller, Revillion, and Bonneterre}{Le~Rhun
  et~al.}{2017}]{Le-Rhun:2017aa}
Le~Rhun, E., Bertrand, N., Dumont, A., Tresch, E., Le~Deley, M.-C., Mailliez,
  A., Preusser, M., Weller, M., Revillion, F., and Bonneterre, J. (2017).
\newblock Identification of single nucleotide polymorphisms of the
  pi3k-akt-mtor pathway as a risk factor of central nervous system metastasis
  in metastatic breast cancer.
\newblock {\em Eur J Cancer} {\bf 87,} 189--198.

\bibitem[\protect\citeauthoryear{Leu, Yan, Fan, Jin, Liu, Curran, Welshons,
  Wei, Davuluri, Plass, Nephew, and Huang}{Leu et~al.}{2004}]{Leu:2004aa}
Leu, Y.-W., Yan, P.~S., Fan, M., Jin, V.~X., Liu, J.~C., Curran, E.~M.,
  Welshons, W.~V., Wei, S.~H., Davuluri, R.~V., Plass, C., Nephew, K.~P., and
  Huang, T. H.-M. (2004).
\newblock Loss of estrogen receptor signaling triggers epigenetic silencing of
  downstream targets in breast cancer.
\newblock {\em Cancer Res} {\bf 64,} 8184--92.

\bibitem[\protect\citeauthoryear{Li and Li}{Li and Li}{2008}]{li2008network}
Li, C. and Li, H. (2008).
\newblock Network-constrained regularization and variable selection for
  analysis of genomic data.
\newblock {\em Bioinformatics} {\bf 24,} 1175--1182.

\bibitem[\protect\citeauthoryear{Li and Li}{Li and Li}{2010}]{li2010variable}
Li, C. and Li, H. (2010).
\newblock Variable selection and regression analysis for graph-structured
  covariates with an application to genomics.
\newblock {\em The annals of applied statistics} {\bf 4,} 1498.

\bibitem[\protect\citeauthoryear{Li and Zhang}{Li and
  Zhang}{2010}]{doi:10.1198/jasa.2010.tm08177}
Li, F. and Zhang, N.~R. (2010).
\newblock Bayesian variable selection in structured high-dimensional covariate
  spaces with applications in genomics.
\newblock {\em Journal of the American Statistical Association} {\bf 105,}
  1202--1214.

\bibitem[\protect\citeauthoryear{Li, Zhang, Wang, Gonzalez, Maresh, Coan,
  et~al\mbox{.}}{Li et~al.}{2015}]{li2015spatial}
Li, F., Zhang, T., Wang, Q., Gonzalez, M.~Z., Maresh, E.~L., Coan, J.~A.,
  et~al. (2015).
\newblock Spatial {B}ayesian variable selection and grouping for
  high-dimensional scalar-on-image regression.
\newblock {\em The Annals of Applied Statistics} {\bf 9,} 687--713.

\bibitem[\protect\citeauthoryear{Li, Yu, Jiang, Shao, Liu, Li, Wu, Zheng, Wu,
  Zhang, Zheng, Qi, Ding, Zhang, and Chang}{Li et~al.}{2018}]{Li:2018aa}
Li, Y.-X., Yu, Z.-W., Jiang, T., Shao, L.-W., Liu, Y., Li, N., Wu, Y.-F.,
  Zheng, C., Wu, X.-Y., Zhang, M., Zheng, D.-F., Qi, X.-L., Ding, M., Zhang,
  J., and Chang, Q. (2018).
\newblock Snca, a novel biomarker for group 4 medulloblastomas, can inhibit
  tumor invasion and induce apoptosis.
\newblock {\em Cancer Sci} {\bf 109,} 1263--1275.

\bibitem[\protect\citeauthoryear{Liu, Chakraborty, Li, Liu, Lozano,
  et~al\mbox{.}}{Liu et~al.}{2014}]{liu2014bayesian}
Liu, F., Chakraborty, S., Li, F., Liu, Y., Lozano, A.~C., et~al. (2014).
\newblock Bayesian regularization via graph laplacian.
\newblock {\em Bayesian Analysis} {\bf 9,} 449--474.

\bibitem[\protect\citeauthoryear{Liu, Chen, Ge, Yan, Huang, Hu, Wen, Li, Huang,
  Qiu, Hao, Yuan, Lei, Yu, and Shao}{Liu et~al.}{2016}]{Liu:2016aa}
Liu, X., Chen, L., Ge, J., Yan, C., Huang, Z., Hu, J., Wen, C., Li, M., Huang,
  D., Qiu, Y., Hao, H., Yuan, R., Lei, J., Yu, X., and Shao, J. (2016).
\newblock The ubiquitin-like protein fat10 stabilizes eef1a1 expression to
  promote tumor proliferation in a complex manner.
\newblock {\em Cancer Res} {\bf 76,} 4897--907.

\bibitem[\protect\citeauthoryear{Lopez, Agoulnik, Zhang, Peterson, Suarez,
  Gandarillas, Frolov, Li, Rajapakshe, Coarfa, Ittmann, Weigel, and
  Agoulnik}{Lopez et~al.}{2016}]{Lopez:2016aa}
Lopez, S.~M., Agoulnik, A.~I., Zhang, M., Peterson, L.~E., Suarez, E.,
  Gandarillas, G.~A., Frolov, A., Li, R., Rajapakshe, K., Coarfa, C., Ittmann,
  M.~M., Weigel, N.~L., and Agoulnik, I.~U. (2016).
\newblock Nuclear receptor corepressor 1 expression and output declines with
  prostate cancer progression.
\newblock {\em Clin Cancer Res} {\bf 22,} 3937--49.

\bibitem[\protect\citeauthoryear{Luo, Pan, and Shen}{Luo
  et~al.}{2012}]{luo2012two}
Luo, C., Pan, W., and Shen, X. (2012).
\newblock A two-step penalized regression method with networked predictors.
\newblock {\em Statistics in biosciences} {\bf 4,} 27--46.

\bibitem[\protect\citeauthoryear{Matthews and Gustafsson}{Matthews and
  Gustafsson}{2006}]{Matthews:2006aa}
Matthews, J. and Gustafsson, J.-A. (2006).
\newblock Estrogen receptor and aryl hydrocarbon receptor signaling pathways.
\newblock {\em Nucl Recept Signal} {\bf 4,} e016.

\bibitem[\protect\citeauthoryear{Nakajima and West}{Nakajima and
  West}{2013a}]{nakajima2013bayesian_a}
Nakajima, J. and West, M. (2013a).
\newblock Bayesian analysis of latent threshold dynamic models.
\newblock {\em Journal of Business \& Economic Statistics} {\bf 31,} 151--164.

\bibitem[\protect\citeauthoryear{Nakajima and West}{Nakajima and
  West}{2013b}]{nakajima2013bayesian_b}
Nakajima, J. and West, M. (2013b).
\newblock Bayesian dynamic factor models: Latent threshold approach.
\newblock {\em Journal of Financial Econometrics} {\bf 11,} 116--153.

\bibitem[\protect\citeauthoryear{Nakajima, West, et~al\mbox{.}}{Nakajima
  et~al.}{2017}]{nakajima2017dynamics}
Nakajima, J., West, M., et~al. (2017).
\newblock Dynamics \& sparsity in latent threshold factor models: A study in
  multivariate eeg signal processing.
\newblock {\em Brazilian Journal of Probability and Statistics} {\bf 31,}
  701--731.

\bibitem[\protect\citeauthoryear{Ni, Stingo, and Baladandayuthapani}{Ni
  et~al.}{2017}]{ni2017bayesian}
Ni, Y., Stingo, F.~C., and Baladandayuthapani, V. (2017).
\newblock Bayesian graphical regression.
\newblock {\em Journal of the American Statistical Association} .

\bibitem[\protect\citeauthoryear{Osborne, Shou, Massarweh, and Schiff}{Osborne
  et~al.}{2005}]{Osborne:2005aa}
Osborne, C.~K., Shou, J., Massarweh, S., and Schiff, R. (2005).
\newblock Crosstalk between estrogen receptor and growth factor receptor
  pathways as a cause for endocrine therapy resistance in breast cancer.
\newblock {\em Clin Cancer Res} {\bf 11,} 865s--70s.

\bibitem[\protect\citeauthoryear{Pan, Xie, and Shen}{Pan
  et~al.}{2010}]{pan2010incorporating}
Pan, W., Xie, B., and Shen, X. (2010).
\newblock Incorporating predictor network in penalized regression with
  application to microarray data.
\newblock {\em Biometrics} {\bf 66,} 474--484.

\bibitem[\protect\citeauthoryear{Park and Casella}{Park and
  Casella}{2008}]{park2008bayesian}
Park, T. and Casella, G. (2008).
\newblock The bayesian lasso.
\newblock {\em Journal of the American Statistical Association} {\bf 103,}
  681--686.

\bibitem[\protect\citeauthoryear{Peng, Zhu, Ander, Zhang, Xue, Sharp, and
  Yang}{Peng et~al.}{2013}]{peng2013integrative}
Peng, B., Zhu, D., Ander, B.~P., Zhang, X., Xue, F., Sharp, F.~R., and Yang, X.
  (2013).
\newblock An integrative framework for bayesian variable selection with
  informative priors for identifying genes and pathways.
\newblock {\em PloS one} {\bf 8,} e67672.

\bibitem[\protect\citeauthoryear{Peng, Eidelberg, and Ma}{Peng
  et~al.}{2014}]{peng2014brain}
Peng, S., Eidelberg, D., and Ma, Y. (2014).
\newblock Brain network markers of abnormal cerebral glucose metabolism and
  blood flow in parkinson?s disease.
\newblock {\em Neuroscience bulletin} {\bf 30,} 823--837.

\bibitem[\protect\citeauthoryear{Peterson, Stingo, and Vannucci}{Peterson
  et~al.}{2016}]{peterson2016joint}
Peterson, C.~B., Stingo, F.~C., and Vannucci, M. (2016).
\newblock Joint bayesian variable and graph selection for regression models
  with network-structured predictors.
\newblock {\em Statistics in medicine} {\bf 35,} 1017--1031.

\bibitem[\protect\citeauthoryear{Polson and Scott}{Polson and
  Scott}{2012}]{polson2012local}
Polson, N.~G. and Scott, J.~G. (2012).
\newblock Local shrinkage rules, l{\'e}vy processes and regularized regression.
\newblock {\em Journal of the Royal Statistical Society: Series B (Statistical
  Methodology)} {\bf 74,} 287--311.

\bibitem[\protect\citeauthoryear{Roberts, Gelman, Gilks, et~al\mbox{.}}{Roberts
  et~al.}{1997}]{roberts1997weak}
Roberts, G.~O., Gelman, A., Gilks, W.~R., et~al. (1997).
\newblock Weak convergence and optimal scaling of random walk metropolis
  algorithms.
\newblock {\em The annals of applied probability} {\bf 7,} 110--120.

\bibitem[\protect\citeauthoryear{Roberts and Rosenthal}{Roberts and
  Rosenthal}{1998}]{roberts1998optimal}
Roberts, G.~O. and Rosenthal, J.~S. (1998).
\newblock Optimal scaling of discrete approximations to langevin diffusions.
\newblock {\em Journal of the Royal Statistical Society: Series B (Statistical
  Methodology)} {\bf 60,} 255--268.

\bibitem[\protect\citeauthoryear{Roberts, Rosenthal, et~al\mbox{.}}{Roberts
  et~al.}{2001}]{roberts2001optimal}
Roberts, G.~O., Rosenthal, J.~S., et~al. (2001).
\newblock Optimal scaling for various metropolis-hastings algorithms.
\newblock {\em Statistical science} {\bf 16,} 351--367.

\bibitem[\protect\citeauthoryear{Schaer, Murphy, and Wolchok}{Schaer
  et~al.}{2012}]{Schaer:2012aa}
Schaer, D.~A., Murphy, J.~T., and Wolchok, J.~D. (2012).
\newblock Modulation of gitr for cancer immunotherapy.
\newblock {\em Curr Opin Immunol} {\bf 24,} 217--24.

\bibitem[\protect\citeauthoryear{Schuster}{Schuster}{2007}]{schuster2007next}
Schuster, S.~C. (2007).
\newblock Next-generation sequencing transforms today's biology.
\newblock {\em Nature methods} {\bf 5,} 16.

\bibitem[\protect\citeauthoryear{Shi and Kang}{Shi and
  Kang}{2015}]{shi2015thresholded}
Shi, R. and Kang, J. (2015).
\newblock Thresholded multiscale gaussian processes with application to
  bayesian feature selection for massive neuroimaging data.
\newblock {\em arXiv preprint arXiv:1504.06074} .

\bibitem[\protect\citeauthoryear{Song and Liang}{Song and
  Liang}{2015}]{song2015split}
Song, Q. and Liang, F. (2015).
\newblock A split-and-merge bayesian variable selection approach for ultrahigh
  dimensional regression.
\newblock {\em Journal of the Royal Statistical Society: Series B (Statistical
  Methodology)} {\bf 77,} 947--972.

\bibitem[\protect\citeauthoryear{Stingo, Chen, Tadesse, and Vannucci}{Stingo
  et~al.}{2011}]{stingo2011incorporating}
Stingo, F.~C., Chen, Y.~A., Tadesse, M.~G., and Vannucci, M. (2011).
\newblock Incorporating biological information into linear models: A bayesian
  approach to the selection of pathways and genes.
\newblock {\em The annals of applied statistics} {\bf 5,}.

\bibitem[\protect\citeauthoryear{Stubelius, Erlandsson, Islander, and
  Carlsten}{Stubelius et~al.}{2014}]{Stubelius:2014aa}
Stubelius, A., Erlandsson, M.~C., Islander, U., and Carlsten, H. (2014).
\newblock Immunomodulation by the estrogen metabolite 2-methoxyestradiol.
\newblock {\em Clin Immunol} {\bf 153,} 40--8.

\bibitem[\protect\citeauthoryear{Tibshirani}{Tibshirani}{1996}]{tibshirani1996regression}
Tibshirani, R. (1996).
\newblock Regression shrinkage and selection via the lasso.
\newblock {\em Journal of the Royal Statistical Society. Series B
  (Methodological)} pages 267--288.

\bibitem[\protect\citeauthoryear{Wolff, Kosyna, Dunst, Jelkmann, and
  Depping}{Wolff et~al.}{2017}]{pmid27823907}
Wolff, M., Kosyna, F.~K., Dunst, J., Jelkmann, W., and Depping, R. (2017).
\newblock Impact of hypoxia inducible factors on estrogen receptor expression
  in breast cancer cells.
\newblock {\em Arch Biochem Biophys} {\bf 613,} 23--30.

\bibitem[\protect\citeauthoryear{Wu, Mao, Li, Yin, Yuan, Chen, Ren, Lu, Li,
  Chen, Chen, Xu, Tian, Lu, Jiang, Zhuang, Chu, and Wu}{Wu
  et~al.}{2018}]{Wu:2018aa}
Wu, S., Mao, L., Li, Y., Yin, Y., Yuan, W., Chen, Y., Ren, W., Lu, X., Li, Y.,
  Chen, L., Chen, B., Xu, W., Tian, T., Lu, Y., Jiang, L., Zhuang, X., Chu, M.,
  and Wu, J. (2018).
\newblock Rage may act as a tumour suppressor to regulate lung cancer
  development.
\newblock {\em Gene} {\bf 651,} 86--93.

\bibitem[\protect\citeauthoryear{Yin, Zhang, Zheng, and Xu}{Yin
  et~al.}{2017}]{Yin:2017aa}
Yin, J., Zhang, Z., Zheng, H., and Xu, L. (2017).
\newblock Irs-2 rs1805097 polymorphism is associated with the decreased risk of
  colorectal cancer.
\newblock {\em Oncotarget} {\bf 8,} 25107--25114.

\bibitem[\protect\citeauthoryear{Yuan, Chen, Lin, Li, Xu, Chen, Hua, and
  Shen}{Yuan et~al.}{2017}]{yuan2017network}
Yuan, X., Chen, J., Lin, Y., Li, Y., Xu, L., Chen, L., Hua, H., and Shen, B.
  (2017).
\newblock Network biomarkers constructed from gene expression and
  protein-protein interaction data for accurate prediction of leukemia.
\newblock {\em Journal of Cancer} {\bf 8,} 278.

\bibitem[\protect\citeauthoryear{Zhang}{Zhang}{2010}]{zhang2010nearly}
Zhang, C.-H. (2010).
\newblock Nearly unbiased variable selection under minimax concave penalty.
\newblock {\em The Annals of statistics} pages 894--942.

\bibitem[\protect\citeauthoryear{Zhang, Jiang, Li, Lv, Li, Qian, Fu, Xu, and
  Guo}{Zhang et~al.}{2015}]{Zhang:2015aa}
Zhang, Y., Jiang, C., Li, H., Lv, F., Li, X., Qian, X., Fu, L., Xu, B., and
  Guo, X. (2015).
\newblock Elevated aurora b expression contributes to chemoresistance and poor
  prognosis in breast cancer.
\newblock {\em Int J Clin Exp Pathol} {\bf 8,} 751--7.

\bibitem[\protect\citeauthoryear{Zhe, Naqvi, Yang, and Qi}{Zhe
  et~al.}{2013}]{zhe2013joint}
Zhe, S., Naqvi, S.~A., Yang, Y., and Qi, Y. (2013).
\newblock Joint network and node selection for pathway-based genomic data
  analysis.
\newblock {\em Bioinformatics} {\bf 29,} 1987--1996.

\bibitem[\protect\citeauthoryear{Zhou and Zheng}{Zhou and
  Zheng}{2013}]{zhou2013bayesian}
Zhou, H. and Zheng, T. (2013).
\newblock Bayesian hierarchical graph-structured model for pathway analysis
  using gene expression data.
\newblock {\em Statistical applications in genetics and molecular biology} {\bf
  12,} 393--412.

\bibitem[\protect\citeauthoryear{Zou}{Zou}{2006}]{zou2006adaptive}
Zou, H. (2006).
\newblock The adaptive lasso and its oracle properties.
\newblock {\em Journal of the American statistical association} {\bf 101,}
  1418--1429.

\bibitem[\protect\citeauthoryear{Zou and Hastie}{Zou and
  Hastie}{2005}]{zou2005regularization}
Zou, H. and Hastie, T. (2005).
\newblock Regularization and variable selection via the elastic net.
\newblock {\em Journal of the Royal Statistical Society: Series B (Statistical
  Methodology)} {\bf 67,} 301--320.

\end{thebibliography}
\end{document}